\documentclass[prd,aps,floatfix,nofootinbib,preprint ,tightenlines,superscriptaddress]{revtex4}
\usepackage{dcolumn}
\usepackage{amsmath}
\usepackage{latexsym}
\usepackage{graphicx}
\usepackage{bm}

\def\svev#1{\left\langle #1\right\rangle}       

\def\Tr{{\rm Tr}\,}

\def\bra#1{\mathinner{\langle{#1}|}}
\def\ket#1{\mathinner{|{#1}\rangle}}

\newcommand{\Lref}{\ensuremath \Lambda_{\rm ref}}

\newcommand{\bee}{\begin{equation}}
\newcommand{\ee}{\end{equation}}
\newcommand{\beea}{\begin{eqnarray}}
\newcommand{\eea}{\end{eqnarray}}

\begin{document}
\title{Repurposing lattice QCD results for composite phenomenology}

\author{Thomas DeGrand}
\email{thomas.degrand@colorado.edu}
\affiliation{Department of Physics, University of Colorado, Boulder, CO 80309, USA}
\author{Ethan T.~Neil}
\email{ethan.neil@colorado.edu}
\affiliation{Department of Physics, University of Colorado, Boulder, CO 80309, USA}

\date{\today}

\begin{abstract}
A number of proposed extensions of the Standard Model include new strongly interacting dynamics, in the form
 of $SU(N)$ gauge fields coupled to various numbers of fermions.  Often, these extensions allow $N=3$ as a
 plausible choice, or even require $N=3$, such as in twin Higgs models, where the new dynamics is a ``copy" of QCD. 
 However, the fermion masses in such a sector are typically different from (often heavier than) the ones of
 real-world QCD, relative to the confinement scale.  Many of the strong interaction masses and matrix 
elements for $SU(3)$ at heavy fermion masses have already been computed on the lattice, typically as a
 byproduct of the approach to the physical point of real QCD.  We provide a summary of these relevant
 results for the phenomenological community.
\end{abstract}
\maketitle

\section{Introduction \label{sec:intro}}

New confining dynamics is a staple of beyond standard model or dark matter phenomenology.
Examples of such systems are                                                             
``hidden valleys'' as a potential source of new physics at colliders \cite{Strassler:2006im,Han:2007ae,Baumgart:2009tn,Pierce:2017taw,Beauchesne:2017yhh},
strongly self-interacting dark matter \cite{Spergel:1999mh} 
which is composite  (Ref.~\cite{Faraggi:2000pv,Cline:2013zca, Boddy:2014qxa,Dienes:2016vei,Berlin:2018tvf,Hochberg:2018rjs} are examples or see \cite{Kribs:2016cew} for a review),           
and even pure gauge systems with non-Abelian symmetry
have a place in
dark matter phenomenology \cite{Faraggi:2000pv,Soni:2016gzf,Acharya:2017szw} or as low energy remnants from the string
landscape \cite{Halverson:2018xge}.

Monte Carlo simulations of lattice regulated quantum field theory can be a resource for such phenomenology
if the model ingredients (gauge fields, scalars, fermions with vector interactions) are favorable.
In some cases the new dynamics is extremely favorable to lattice simulation -- it involves
$SU(3)$ gauge dynamics and (nearly) degenerate flavors of fundamental representation fermions.
A particularly natural example of such systems occurs in ``twin Higgs'' or mirror-matter models, in which the choice of $SU(3)$ is required
(a partial set of citations are
Refs.~\cite{Chacko:2005pe,Barbieri:2015lqa,Garcia:2015loa,Garcia:2015toa,Craig:2015xla,Cheng:2015buv,Chacko:2015fbc,Craig:2016kue,Chacko:2018vss,Hochberg:2018vdo,Kilic:2018sew,Terning:2019hgj,Chacko:2019jgi,Harigaya:2019shz}).  Other models may be more general but include $SU(3)$ with fundamental representation fermions as a possibility (for example, Refs.~\cite{Kilic:2009mi,Bai:2010qg,Bai:2013xga,Antipin:2015xia,Harigaya:2015ezk,Curtin:2015jcv,Harigaya:2016pnu,Batell:2017kho}).

In many cases, the fermion masses needed to carry out the phenomenologist's task do not coincide
with those of the real world up, down, strange $\dots$ quark masses; they are heavier than these
physical values. (We will define more precisely what we mean by ``heavier,'' below.)
Lattice practitioners have studied these systems for many years.
This is because the cost of QCD simulations scales as a large inverse power of the pion mass.
Simulations ``at the physical point'' (where $M_\pi\sim 140$ MeV) are a relatively
recent development. However, results from these heavier mass systems
are usually presented as not being interesting on their own; they are simply intermediate results on the way
to the physical point. This means that they are sometimes not presented in a way which is accessible
to researchers outside the lattice community.

The purpose of this manuscript is to provide an overview of QCD lattice results away from the physical point of QCD,
which can impact beyond standard model (BSM) phenomenology. We will try to do this in a way which is useful to
physicists working in this area (rather than to researchers doing lattice gauge theory; we have previously written another paper on this subject aimed at the lattice community \cite{DeGrand:2018sao}).
Most of the data we will show is taken from the lattice literature. Some of it is our own.
When we show our own data, we do not intend that it be taken as having higher quality than what 
might be elsewhere in the literature, only that we could not easily find precisely what we wanted to present.
Most of what we are showing is generic lattice data.

Our focus here is on QCD with moderately heavy quarks, where ``moderately'' means that the quarks are not so heavy that
they are no longer important for the dynamics of the theory.  For sufficiently heavy quarks, the dynamics
becomes that of a pure Yang-Mills gauge theory.  We do not present numerical results for pure-gauge theory here,
 but instead direct the interested reader to a review of the extensive lattice literature on
the large-$N_c$ limit of pure-gauge SU$(N_c)$ \cite{Lucini:2012gg}.

The outline of the paper is as follows: We make  some brief remarks about lattice simulations aimed at
phenomenologists (Sec.~\ref{sec:intro_ph}).
Then we describe hadron spectroscopy in Sec.~\ref{sec:spectro}, including spectroscopy of pseudoscalar 
mesons in Sec.~\ref{sec:pions}, other 
 mesons and baryons in Sec.~\ref{sec:meson_baryon}, and other states in Sec.~\ref{sec:other}.
We describe results for vacuum transition matrix elements (i.~e.~decay constants) in Sec.~\ref{sec:matel}.
We then turn to strong decays, 
describing $\rho \rightarrow \pi\pi$ in Sec.~\ref{sec:decay} and to the mass and width of the
$f_0$ or $\sigma$ meson in subsection \ref{sec:sigma}.
Our conclusions are found in Sec.~\ref{sec:conc}.
The appendices contain technical details relevant to our own lattice simulations.

\section{Remarks about lattice QCD for BSM phenomenologists \label{sec:intro_ph}}
In our experience, the approaches taken by
lattice practitioners and beyond standard model phenomenologists toward the theories they study
are somewhat different. To the lattice practitioner the confining gauge dynamics is paramount
and everything else is secondary. Flavor standard model quantum numbers of the constituents
generally play little role in a lattice simulation, as do their electroweak interactions.

We illustrate the difference in approaches by highlighting a few key points where the phenomenologist 
and the lattice simulator are most likely to have a different picture of the same physics.

\subsection{Perturbative interactions are treated separately}

Lattice simulations are conducted at a finite lattice spacing $a$ and with a finite number of 
sites $N_s$ (so they are done in a finite ``box'' with length $L = N_s a$.)  As a result, only the range of scales 
between the infrared cutoff $\Lambda_{IR} \sim 1/L$ and the ultraviolet cutoff $\Lambda_{UV} \sim 1/a$ can
 be treated fully dynamically.  State-of-the-art QCD simulations will have roughly a factor of 100 
separating the two cutoffs, for example placing the boundaries at $1/L \sim 50$ MeV and $1/a \sim 5$ GeV;
 this is plenty of room for confinement physics, but cannot accommodate the electroweak scale of the
 standard model directly \footnote{There is an additional technical problem, which is that chiral gauge 
theories cannot be treated with standard lattice 
methods \cite{Nielsen:1980rz,Nielsen:1981xu,Nielsen:1981hk} - see \cite{Kaplan:2009yg} for a 
contemporary review.  We are fortunate that the electroweak scale is well-separated from the QCD scale in the real world.}.

This is not a problem for simulating the standard model, simply because the electroweak interactions 
are perturbative around the QCD confinement scale.  (QED is treated in the same way to zeroth order, in fact.)
  The idea of \emph{factorization}, crucial to perturbative treatments of jet physics and other aspects of QCD,
 allows the treatment of non-perturbative effects by calculating QCD matrix elements in isolation.  For example,
 an electroweak decay of hadronic initial state $\ket{i}$ mediated by short-distance operator $\mathcal{O}$ can
 be factorized into the purely hadronic transition matrix element $\bra{f} \mathcal{O} \ket{i}$, times electroweak
 and kinematic terms.  (This is a simplified story: accounting for momentum dependence and contributions from 
multiple operators to the same physical process can lead to complicated-looking formulas in terms of multiple 
form factors.  But the basic idea is the same.)

As a simple but concrete example, consider the electroweak decay of the pseudoscalar charm-light
 meson $D \rightarrow \ell \nu$.  The partial decay width for this process is given by \cite{Tanabashi:2018oca}
\bee
\Gamma(D \rightarrow \ell \nu) = \frac{M_D}{8\pi} f_D^2 G_F^2 |V_{cd}|^2 m_\ell^2 \left(1 - \frac{m_\ell^2}{M_D^2} \right)^2.
\ee
Here $G_F^2 |V_{cd}|^2$ are the electroweak couplings, the masses of the $D$ meson and lepton appear due to 
kinematics, and the strongly-coupled QCD physics is contained entirely in the decay constant $f_D$, which 
is proportional to the matrix element $\bra{0} \mathcal{A}^\mu \ket{D}$ giving the overlap of the initial 
state $D$ through the axial vector current with the final state (the vacuum, since from the perspective of
 QCD there is nothing left in the final state.)  A lattice calculation of $f_D$ is a necessary input to
 predicting this decay rate in the standard model.  Conversely, knowing $f_D$ allows one to determine the 
electroweak coupling $|V_{cd}|$ from the observed decay width.

In particular, working in the low-energy effective theory means that the Yukawa couplings of the quarks to the 
Higgs boson never appear explicitly; instead, the Higgs is integrated out and only vector-like mass terms of
 the form $m_q \bar{q} q$ are included.  These quark masses, along with the overall energy scale of the 
theory $\Lambda$, are the only continuous free parameters of the strongly-coupled theory in isolation.

\subsection{Lattice simulations produce dimensionless ratios of physical scales}

The ingredients of a lattice calculation are a set of bare (renormalizable) couplings,
 a dimensionless gauge coupling and
a set of (dimensionful) fermion masses, a UV cutoff (the lattice spacing), 
 and (implicitly) a whole set of irrelevant operators arising from the
particular choice made when the continuum action is discretized. 

All lattice predictions are of dimensionless 
quantities; for example, a hadron mass $m$ will
be determined as the dimensionless product $am$, where $a$ is the lattice spacing.  If a
mass appears alone in a lattice paper, the authors are likely working in ``lattice units'' where the $a$ is
included implicitly.
Taking the continuum limit involves tuning the bare parameters so that correlation lengths measured in units
of the lattice spacing diverge: in this limit the UV cutoff becomes large with respect to other dimensionful
parameters in the theory.

All lattice predictions are functions of $a$; only their  $a\rightarrow 0$ values are physical.
Often, these dimensionless quantities are ratios of dimensionful ones, such as mass ratios.
In an asymptotically free theory, if the lattice spacing is small enough, any
dimensionless quantity, such as a mass ratio, will behave as
\bee
[a m_1 (a)]/[a m_2 (a)]  = m_1(0)/m_2(0) + {\cal O}(m_1a) +  {\cal O}[(m_1 a)^2] +\dots,
\label{eq:scaling}
\ee
modulo powers of $\log(m_1a)$.
The leading term  is the cutoff-independent prediction.  Everything else is an artifact
of the calculation.

To use equation~(\ref{eq:scaling}) to make a prediction for a dimensionful quantity (like a mass, $m_1$) requires choosing 
some fiducial $(m_2)$ to set a scale. Lattice groups make many different choices for reference scales, 
mostly based on ease of computation (since an uncertainty in the scale is part of the error budget for any
lattice prediction). Masses of particles (the rho meson, the $\Omega^-$) or leptonic decay constants 
such as $f_\pi$ or $f_K$ are simple and intuitive reference scales, but are not always the most numerically
 precise.  Other common choices include inflection points on the heavy quark potential
 (``Sommer parameters''  $r_0$ and $r_1$~\cite{Sommer:1993ce})
 or more esoteric quantities derived from
the behavior of the  gauge action under some smoothing scheme (``gradient flow'' or
 ``Wilson flow'' \cite{Luscher:2010iy,other}, some of the corresponding length scales 
are $\sqrt{t_0}$, $\sqrt{t_1}$, and $w_0$).  These latter choices, which may be thought of 
roughly as setting the scale using the running of the gauge coupling constant, are computationally
 inexpensive and precise, but their values in physical units must be determined by matching on to experiment
 in other lattice calculations.  We reproduce here approximate current values for these reference scales in the continuum limit:
 \beea
 r_0 &\approx& 0.466(4)\ \rm{fm} \\
 r_1 &\approx& 0.313(3)\ \rm{fm} \\
 \sqrt{t_0} &\approx& 0.1465(25)\ \rm{fm} \\
 w_0 &\approx& 0.1755(18)\ \rm{fm}
 \eea
taken from \cite{Davies:2009tsa} (${r_0}$, ${r_1}$) and \cite{Borsanyi:2012zs} ($\sqrt{t_0}$, ${w_0}$). 

For the purposes of BSM physics, the appropriate choice of physical state for scale setting is likely to depend on 
what sort of model one is considering.  For a composite dark matter model, the mass of the dark matter candidate
baryon or meson is often a natural choice.  In the context of composite Higgs models, the physical Higgs vev is often related
closely to the decay constant of a ``pion''.  Ultimately, the choice is a matter of convenience; we can always
 exchange dimensionless
ratios with one scale for ratios with another scale.  But some caution is required if these ratios are taken 
at finite $a$, since the additional
artifact terms in Eq.~(\ref{eq:scaling}) may be different.

Some phenomenological, qualitative discussions of QCD like to set the units in terms 
of a ``confinement scale'', $\Lambda_{\rm QCD}$.
There is no such physical scale.  Plausible choices for a physical scale associated with confinement 
could include the proton mass $\sim$ 1 GeV,
the rho meson mass $\sim 800$ MeV, the breakdown scale of chiral perturbation theory
 $4\pi F_\pi \sim 1.6$ GeV, and many other options.  In some cases, $\Lambda_{\rm QCD}$ may
 refer to the perturbative $\Lambda$ parameter, also known as the dimensional transmutation
 parameter, defined as an integration constant of the perturbative running coupling; its 
value in the $\overline{MS}$ scheme is a few hundred MeV, depending on the number of quark 
flavors included \cite{Tanabashi:2018oca}.  This is a particularly awkward choice to use in 
conjunction with lattice results, since it is not renormalization-scheme independent.  

In certain cases, it may be desirable to work in terms of the strong coupling constant, which is specified at some high energy scale and then run down to low energies.  There are some lattice calculations of $\Lambda_{\overline{MS}}$ available \cite{Gockeler:2005rv,Blossier:2011tf}, which can be used to match on to the perturbative $\Lambda$ parameter determined in such a running coupling calculation as a starting point for use of other lattice results.  However, we strongly recommend the use of more physical reference scales whenever possible, and we further recommend replacing calculations of physical processes that involve the strong coupling constant with lattice matrix elements, using the idea of factorization as discussed in the previous subsection.

\subsection{Quark masses are inconvenient free parameters}

From the perspective of QCD as a quantum field theory, the quark masses are completely free parameters. 
 For a particular lattice simulation, the quark masses are also free parameters, but they must be fixed 
as inputs in the form $am_q$ before the simulation is run.  This means that the dimensionless ratios of 
each $m_q$ to our chosen reference scale $\Lref$ are not adjustable without starting a new lattice simulation. 
 Extrapolation can be done to approach the massless limit $m_q / \Lref \rightarrow 0$ or the pure-gauge theory
 limit $m_q / \Lref \rightarrow \infty$.
If results are desired from some nonzero value of a quark mass, there is a further
 tuning involving ratios of the fermion masses both among themselves
and with respect to some overall energy scale.

Numerical values of the form $m_q / \Lref$ almost never appear in lattice papers, because the quark masses 
themselves suffer from the same issue as $\Lambda_{\overline{MS}}$: they are not renormalization-group invariant.
  One can extract results for quark masses in a particular renormalization scheme like $\overline{MS}$ from 
lattice calculations, but it requires careful perturbative matching and is not done as a matter of course in
 most lattice QCD work.  Instead, it is common practice to use another physical observable (typically a hadron
 mass) as a proxy for the quark mass.

In principle, any physical observable which depends directly on quark mass can be a good proxy; the best
 proxies depend strongly on the quark mass.  A common convention in lattice QCD is to use the squared pion
 mass to fix the light-quark mass, relying on the Gell-Mann-Oakes-Renner (GMOR)
 relation \cite{GellMann:1968rz,Gasser:1983yg}, $M_\pi^2 = 2 \langle \bar{q} q  \rangle m_q$.  
(This leads to the common shorthand amongst lattice practitioners of asking ``How heavy are your pions?''
 to judge the approximate masses of the light quarks in a given study.)  We will adopt this approach and 
elaborate on it in Sec.~\ref{sec:spectro} below.  Masses of heavy quarks (charm, bottom) must be matched 
to corresponding hadronic states containing valence charm and bottom quarks.

To use lattice QCD results in some new physics scenario, one would introduce a new reference scale
$ \Lref'$ and fix any fermion masses $m_q' / \Lref'$, most likely using a proxy as discussed above. Then any matrix element which had an energy scaling exponent $p$ would simply be related to the QCD result at the same ratio of fermion mass to reference scale,
\bee
\svev{O'(r_q)} = \svev{O_{QCD}(r_q)}\left(\frac{\Lref'}{\Lref}\right)^p,
\ee
where $r_q = m_q / \Lref = m_q' / \Lref'$.  For example, if lattice QCD simulations with $(M_\pi / M_\rho)^2 \sim 0.8$ give a vector meson mass of 1.5 GeV and a nucleon mass of 2.3 GeV, then for a composite dark matter model with a dark nucleon mass of 1 TeV, using the nucleon mass as a reference scale the corresponding dark vector meson mass would be 650 GeV at the same proxy ratio $(M_P / M_V)^2 \sim 0.8$.

\subsection{Changing the number of flavors and colors is somewhat predictable}

Most lattice simulations involve two flavors of degenerate light fermions, emulating the up and down quarks.
QCD plus QED simulations are beginning to break this degeneracy. Many simulations include a strange quark
at its physical mass value and some simulations also include the charm quark.
 Little work has been done on QCD with 
a single light flavor \cite{Farchioni:2007dw}
or for systems with a large hierarchy between the up and down quark mass.
There is also little lattice literature about systems  with non-fundamental representations of fermions.

The Flavour Lattice Averaging Group (FLAG) \cite{Aoki:2019cca} says
``In most cases, there is reasonable agreement among results with $N_f=2$, 2+1, and 2+1+1'' 
for
fermion masses, low energy chiral constants, decay constants, 
the QCD $\Lambda$ parameter,
and the QCD running coupling measured at the $Z$ pole. This is actually a restricted statement: $N_f$ values
range over two degenerate light quarks, plus a strange quark at around its real world quark mass, plus
a charm quark near the physical charm mass. As far as we can tell from looking at simulations, results for
QCD with up to 4-6 light degenerate flavors are not too different from ``physical'' QCD. This begins to
break down as $N_f$ rises, and by $N_f=8$ the spectrum is not very QCD like. At some point $SU(3)$ systems
cross over from confining into infrared conformal behavior, which is definitely different 
from QCD (see the reviews Ref.~\cite{DeGrand:2015zxa,Svetitsky:2017xqk}).

Confining systems in beyond standard model phenomenology are not restricted to $N=3$, of course. Lattice results for
such systems are much more sparse than for $N=3$. We will not describe lattice results away from $N=3$,
other than to say that 't Hooft scaling seems to work well as a zeroth order description of masses and
of the studied matrix 
elements \cite{DeGrand:2012hd, Bali:2013kia, DeGrand:2013nna, Cordon:2014sda, DeGrand:2016pur, Hernandez:2019qed}. 
 As is well known, fermion loops essentially decouple from the theory in the 't Hooft large-N limit, so that 
lattice QCD results for pure Yang-Mills gauge theory (also known as the ``quenched'' limit in lattice literature) 
are likely to be relevant: see \cite{Lucini:2012gg} for a review.


\section{Spectroscopy \label{sec:spectro}}

The spectroscopy of the lightest flavor non-singlet meson states and of the lightest baryons (the
octet and decuplet) at the physical point is basically a solved problem.
Lattice calculations reproduce experimental data at the few per cent level. These calculations
carefully take into account extrapolations to zero lattice spacing, infinite volume,
and to the physical values of the light quarks.  We refer readers to the literature
and just take the known values of real-world hadron masses as given when we use them in comparisons.
Flavor singlet states often have significant overlap with intermediate strong scattering states, and must be treated
carefully; we will discuss results for one such state, the $f_0$ or $\sigma$ meson, in Sec.~\ref{sec:sigma} below.

Spectroscopy at unphysical (heavier) quark masses 
is not so well studied, but published data is probably accurate at the five to ten 
percent level. This should be sufficient for composite model building.
Our pictures and analysis are based on three primary sources and one secondary one.
Ref.~\cite{WalkerLoud:2008bp} used a $20^3\times 64$ site lattice at a lattice 
spacing $a=0.12406$ fm.
Ref.~\cite{Aoki:2008sm} performed simulations on a $32^3\times 64$ lattice at a lattice 
spacing of $a=0.0907$ fm.  These sources both include two degenerate light quarks and a strange quark.
Refs.~\cite{Alexandrou:2008tn,Jansen:2009hr,Baron:2009wt} include results for $24^3 \times 48$
 and $32^3 \times 64$ lattices, with approximate lattice spacings of $a=0.09$ and $0.07$ fm. 
 (We include data only from the set of ensembles labeled `B2'-`B6' and `C1'-`C4', for which we were
 able to find the majority of the results we are interested in.)  This set of simulations omits the 
strange quark, including only two degenerate light quarks.

Finally, we add our own smaller lattice simulations with two degenerate fermions on a 
$16^3\times 32$ lattice, at a lattice
spacing of about $0.1$ fm, an extension of a set of simulations performed by one of us \cite{DeGrand:2016pur}.
  This set is compromised by its small volume, although such effects are mitigated
by the large quark masses, and we will see that they are qualitatively similar to
the other data sets where they overlap.  These new results extend to much heavier quark masses than the primary sources.

All these simulations are at fixed lattice spacing; in the plots we take the published lattice
spacings and convert the lattice data to physical units from it.  Based on experience with lattice
QCD, we believe the systematic error due to lack of continuum extrapolation is unlikely to be
larger than five to ten percent at the lattice spacings given here.  The generally close agreement of the 
$a \sim 0.12$ fm, $a \sim 0.09$ fm and $a \sim 0.07$ fm results to this level of accuracy reinforces this expectation.  

We remark (again) that these are merely representative lattice data sets, which 
we found presented in an easy to use format.  All of the results we will show include quarks which
 are much heavier than the physical up and down quarks, but still light compared to the confinement scale.  

To emphasize that the various hadronic states we will be studying do not take on their real-world 
QCD masses in these simulations, we will label states by spin and parity: for example, we will denote 
the lightest pseudoscalar meson ($\pi$ in QCD) as ``PS'', the vector and axial-vector mesons as ``V'' and ``A'', etc.

\subsection{Pseudoscalar mesons and setting the quark mass\label{sec:pions}}

\begin{figure}
\begin{center}
\includegraphics[width=0.8\textwidth,clip]{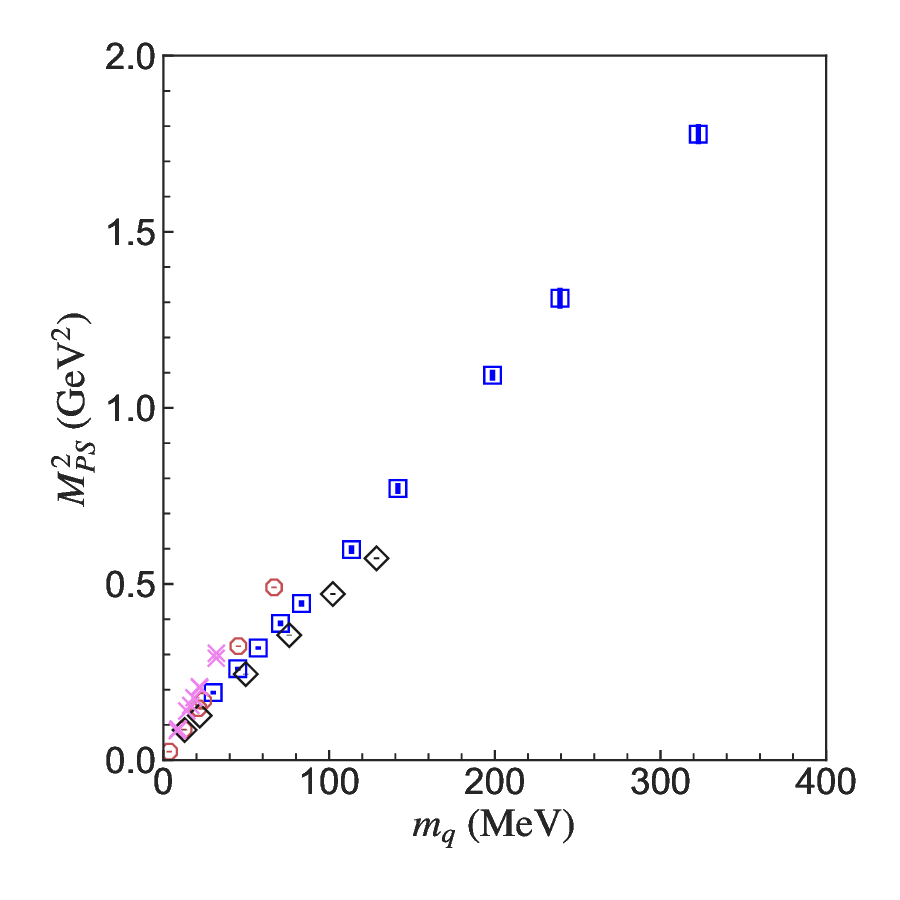}
\end{center}
\caption{Squared pseudoscalar meson mass in GeV${}^2$ as a function of the quark mass in MeV.
Data are black diamonds from Ref.~{\protect{\cite{WalkerLoud:2008bp}}},
red octagons from Ref.~{\protect{\cite{Aoki:2008sm}}},
violet crosses from Refs.~{\protect{\cite{Alexandrou:2008tn,Jansen:2009hr,Baron:2009wt}}},
blue squares from this work.  The square points are likely contaminated by finite-volume systematic effects, 
as discussed in the text, but they nevertheless show the correct qualitative relationship between $M_{PS}^2$ and $m_q$.
\label{fig:mpi2mq}}
\end{figure}

We begin with the pseudoscalar light-quark mesons, i.e. the pions, which are the lightest states in the 
spectrum and exhibit special quark mass behavior due to their nature as pseudo-Goldstone 
bosons: $M_{PS}^2 \sim m_q$.  This approximately linear behavior, shown in Fig.~\ref{fig:mpi2mq}, 
is found to persist even out to large quark mass.  A plot of $M_{PS}^2 / m_q$ would reveal the (small) 
curvature in the data, which would be described well by chiral perturbation theory.  Note that the quark 
masses shown here are not consistent in terms of renormalization scheme: some values are $\overline{MS}$ 
and others are in the lattice scheme.  As a result, we caution that Fig.~\ref{fig:mpi2mq} should only be 
taken as a qualitative result.

Due to the issues of renormalization scale and scheme dependence, the quark mass itself is not the most useful 
variable to present results against, as discussed in Sec.~\ref{sec:intro_ph}.  One alternative would be to only
 make plots of dimensionless ratios; with one free dimensionless parameter in this
 case, $m_q / \Lambda_{\rm ref}$, plotting two dimensionless ratios should show a single 
universal curve as the quark mass is varied.  Such a global picture, known as an ``Edinburgh plot'', often 
appears in exploratory lattice publications.  

\begin{figure}
\begin{center}
\includegraphics[width=0.8\textwidth,clip]{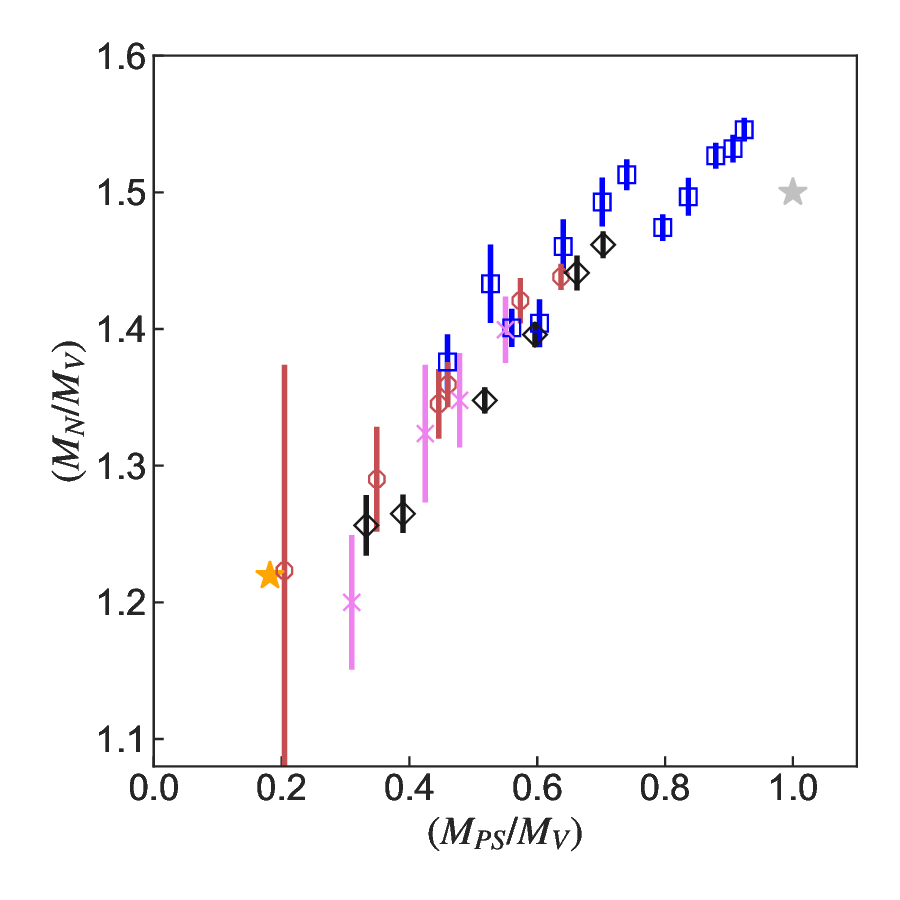}
\end{center}
\caption{
Edinburgh plot, 
$M_N/M_V$ vs $M_{PS}/M_V$.
Data are black diamonds from Ref.~{\protect{\cite{WalkerLoud:2008bp}}},
red octagons from Ref.~{\protect{\cite{Aoki:2008sm}}},
violet crosses from Refs.~{\protect{\cite{Alexandrou:2008tn,Jansen:2009hr,Baron:2009wt}}},
blue squares from this work.  The stars show the physical point and the heavy quark limit.
\label{fig:ed}}
\end{figure}

Fig.~\ref{fig:ed} shows an Edinburgh plot of the nucleon to vector mass ratio versus 
the pseudoscalar to vector meson mass ratio.  This curve is, of course, independent of the 
overall confinement scale; it captures the behavior of QCD as the fermion mass is taken from
small (or zero) to infinity.  Other systems (different gauge groups, different fermion composition) would 
have their own Edinburgh plots with different curves.

\begin{figure}
\begin{center}
\includegraphics[width=0.65\textwidth,clip]{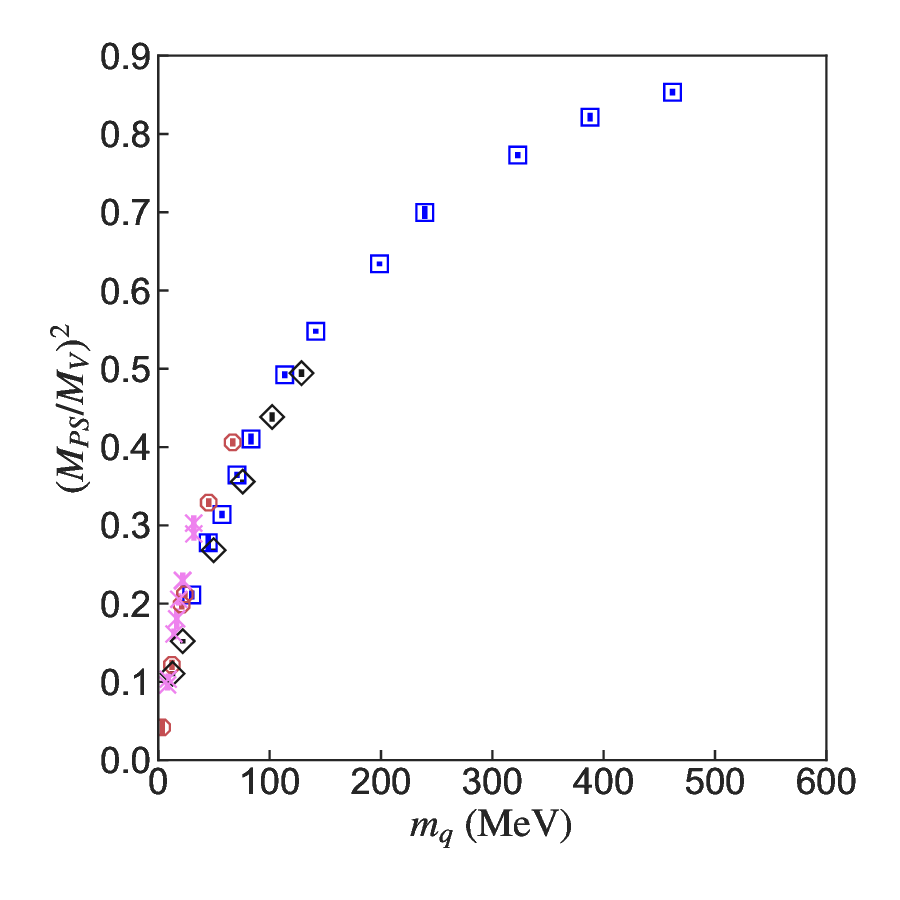}
\includegraphics[width=0.65\textwidth,clip]{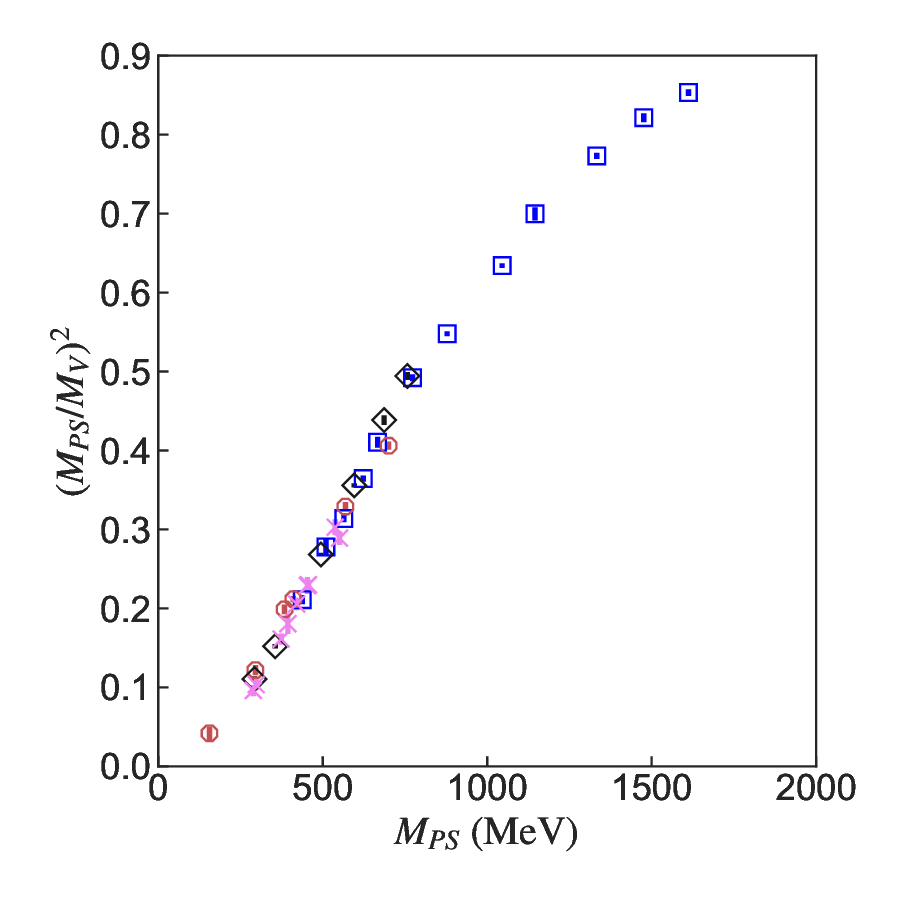}
\end{center}
\caption{Top panel: the ratio $(M_{PS}/M_V)^2$ as a function of the quark mass in MeV.
Data are black diamonds from Ref.~{\protect{\cite{WalkerLoud:2008bp}}},
red octagons from Ref.~{\protect{\cite{Aoki:2008sm}}},
violet crosses from Refs.~{\protect{\cite{Alexandrou:2008tn,Jansen:2009hr,Baron:2009wt}}},
blue squares from this work.  Bottom panel: the ratio $(M_{PS}/M_V)^2$ as a function of $M_{PS}$ in MeV.
\label{fig:pirho}}
\end{figure}

Although the Edinburgh plot has some nice features, in particular capturing the full 
variation of $m_q / \Lambda_{\rm ref}$ from zero to infinity in a finite range, it obscures 
the detailed parametric dependence of individual quantities on quark mass.  As such, it is not a 
convenient way to use results for application to phenomenological models.  To present
results in a way which is easier to work with, we will instead choose the variable $(M_{PS}^2 / M_V^2)$ as
 a proxy for $m_q$.  This is a dimensionless quantity running from zero (at zero quark mass) to unity
 (in the heavy quark limit), which is also (roughly) linear in the quark mass at small $m_q$.

Panel (a) of Fig.~\ref{fig:pirho} shows this ratio as a function of the quark mass.  The mass dependence 
of $M_V$ itself spoils the linear dependence of $M_{PS}^2 / M_V^2$ on $m_q$, but it is still monotonic, 
making this a reasonable replacement for $m_q$ as a parameter.  Much of the data we want to quote is 
presented as being generated at some pion mass
in MeV. A translation plot between $M_{PS}$ and $(M_{PS}/M_V)^2$ is shown in panel (b).  We will present 
results going forward exclusively using $(M_{PS}/M_V)^2$, but these conversion figures may be useful in 
translating from other sources.  Inspection of panel (b) shows that up to approximately 
$M_{PS} \sim 1$ GeV, $(M_{PS} / M_V)^2$ is roughly linear in $M_{PS}$; fitting to the data in this range gives the relation
\bee \label{eq:mPmVsq_vs_mP}
\left(\frac{M_{PS}}{M_V}\right)^2 \approx -0.11 + 0.77 \frac{M_{PS}}{1\ \rm{GeV}},
\ee
which is accurate to within a few percent over the range of data considered, 
$200\ {\rm MeV} \lesssim M_{PS} \lesssim 1000\ {\rm MeV}$ (or equivalently, $0.1 \lesssim (M_{PS} / M_V)^2 \lesssim 0.7$.)
  This is completely empirical, and in particular must break down for sufficiently light quark masses; in the limit
 of zero $M_{PS}$, $M_V$ will become approximately constant and we should recover quadratic dependence of $(M_{PS} / M_V)^2$ on $M_{PS}$.

To briefly summarize our treatment of quark-mass dependence:

\begin{itemize}

\item In the intermediate quark-mass regime $0.1 \lesssim (M_{PS} / M_V)^2 \lesssim 0.7$ (roughly equivalent
 to $200\ {\rm MeV} \lesssim M_{PS} \lesssim 1000\ {\rm MeV}$ or $20\ {\rm MeV} \lesssim m_q \lesssim 300\ {\rm MeV})$,
 the quantity $(M_{PS} / M_V)^2$ is roughly linear in $M_{PS}$, following Eq.~(\ref{eq:mPmVsq_vs_mP}).  Other quantities
 will also show simple linear dependence on $(M_{PS} / M_V)^2$.  This regime is our main focus in this paper.
\item At lighter quark masses, there is a qualitative change in dependence on $(M_{PS} / M_V)^2$ for many quantities. 
 In this regime, one may rely on experimental results for real-world QCD or effective theories such as chiral perturbation theory.
\item At heavy quark masses, there is also a qualitative change in the dependence on $(M_{PS} / M_V)^2$. 
 Here we will find that returning to the use of the quark mass itself as a parameter is the best way to describe the data. 
\end{itemize}

We will attempt to make some connection to this final heavy-quark regime in what follows, but we caution 
the reader that lattice results may be particularly unreliable here, as large and potentially uncontrolled 
lattice artifacts are expected to appear as $am_q \rightarrow 1$.  There are some reliable lattice data 
for quarkonia which could be applicable that make use of fine lattice spacings, high precision, and/or 
specialized lattice actions to overcome the discretization effects.  Unfortunately, all of the lattice 
data we have found for quarkonium systems is specific to the charm and bottom-quark masses, requiring model 
extrapolation for more general use.  A more general study of quarkonium properties could be an interesting future lattice project.

\subsection{Other mesons and baryons\label{sec:meson_baryon}}

We begin with the other pseudoscalars, which are the next lowest-lying states above the pion.  The $K$ 
and $\eta$ are Goldstone bosons associated with breaking of $SU(3) \times SU(3)$ flavor symmetry including
 the strange quark: given a value for the light-quark mass $m_q$ and a strange quark mass $m_s$, their
 masses are predicted by chiral perturbation theory at leading order to follow the GMOR relation
\beea
M_K^2 &=&  \frac{m_q+m_s}{2m_q} M_\pi^2, \\
M_\eta^2 &=&  \frac{m_q + 2m_s}{3m_q} M_\pi^2.
\eea
At very heavy quark masses, these formulas will break down.  We further emphasize that these states are 
distinct from the pions only by virtue of including a strange quark with $m_s \neq m_q$.  For application
 to a new physics model where there are only two light quarks, the $K$ and $\eta$ do not exist as distinct meson states.

The $\eta'$ meson is a special case; it is much heavier than the other pseudoscalar mesons due to 
the influence of the $U(1)_A$ anomaly.  An example lattice QCD calculation of this state 
is \cite{Michael:2013vba}; they find weak dependence on the light-quark masses, similar to the $\eta$ meson.   The relation 
\bee
M_{\eta'} / M_{\eta} \approx 1.8(0.2)
\ee
from the reference is fairly accurate over a wide range of $M_{PS}$.

\begin{figure}
\begin{center}
\includegraphics[width=0.7\textwidth,clip]{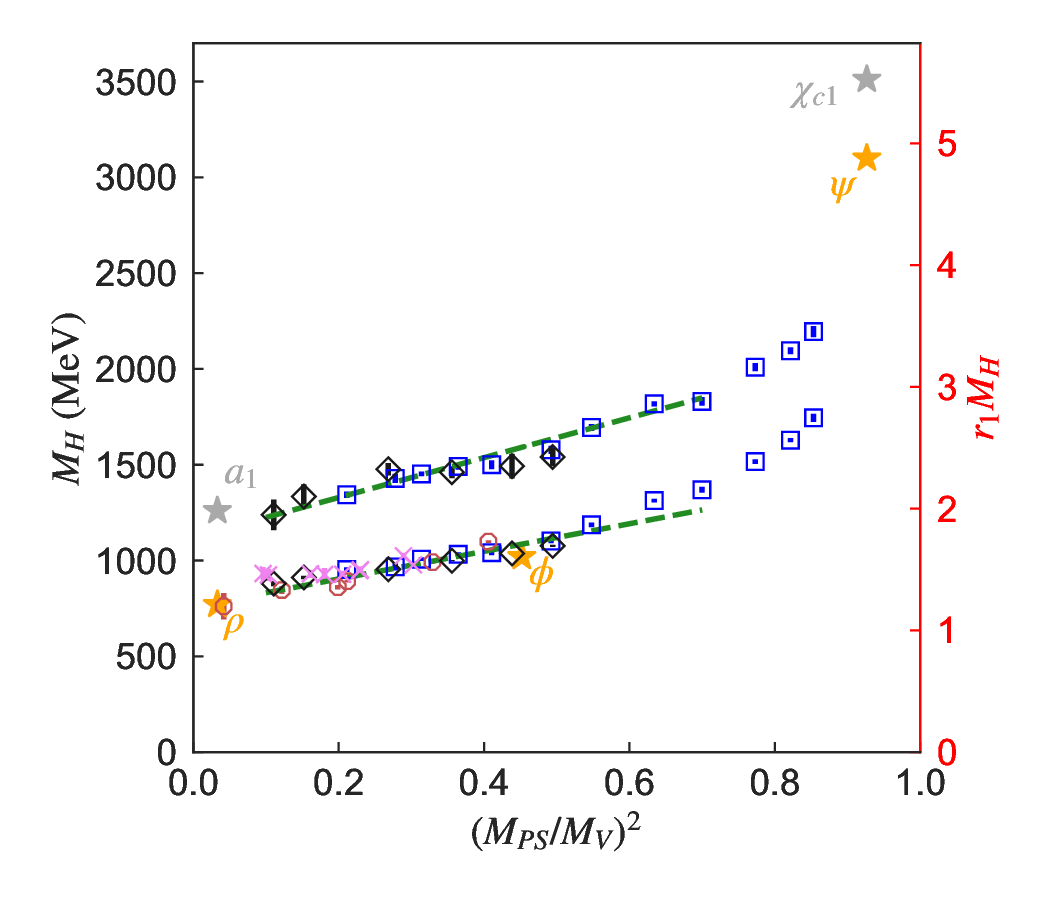}
\includegraphics[width=0.7\textwidth,clip]{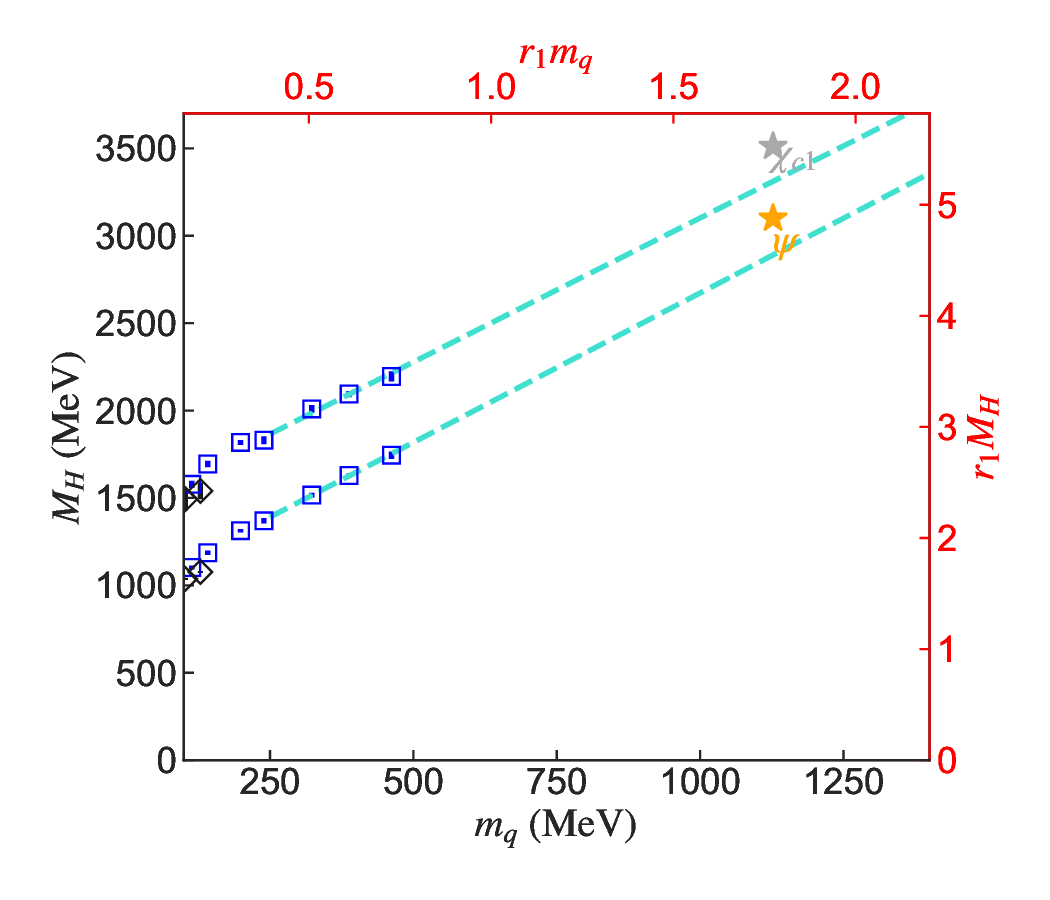}
\end{center}
\caption{Meson masses in MeV as a function of the  ratio $(M_{PS}/M_V)^2$ (top panel) and quark mass in MeV (bottom panel.) 
Stars are values of physical particles, obtained as described in the text: gold (silver) stars denote vector (axial-vector) states.
The lower densely-populated band is the mass of the isovector vector meson (the rho)
and the upper band is the $a_1$. For these particles, the symbols are black diamonds from Ref.~{\protect{\cite{WalkerLoud:2008bp}}},
red octagons from Ref.~{\protect{\cite{Aoki:2008sm}}},
violet crosses from Refs.~{\protect{\cite{Alexandrou:2008tn,Jansen:2009hr,Baron:2009wt}}},
blue squares from this work.  The dashed lines show linear fits to the data in certain regimes, as described in the text.
\label{fig:meson}}
\end{figure}

The next lightest states commonly reported in lattice simulations are the vector mesons $\rho$
 (isospin-triplet) and $\omega$ (isospin-singlet), and the axial-vector meson $a_1$.  These states 
are easily isolated as ground states from correlation functions with the corresponding symmetry 
properties.  Fig.~\ref{fig:meson} (top panel) shows various vector-meson masses as a function of 
$(M_{PS}/M_V)^2$. We have added the physical values of the $\rho$ and $a_1$ mesons to the plot.
We have also added the phi ($\bar{s} s$ vector) meson. For it, we need a corresponding pseudoscalar mass; we use the
GMOR relation to define a ``strange eta'' or $s\bar s$ pseudoscalar, in the absence of
$\eta-\eta'$ mixing, using $M_{\eta_s}^2=2m_K^2-m_\pi^2$. Its mass is 685 MeV, giving a mass 
ratio $(M_{PS}/M_V)^2 = 0.45.$  Finally, we include the vector $J/\psi$ and axial-vector $\chi_{c1}$ 
charmonium states, taking the $\eta_c$ as the corresponding pseudoscalar (which yields $(M_{PS} / M_V)^2 = 0.93$.)

A useful way to present the results in Fig.~\ref{fig:meson} is to provide a simple linear
parameterization of each mass as a function of $x=(M_{PS}/M_V)^2$,  
\bee
M_H = A_H + B_H x .
\ee
As discussed above, and as is evident from the plot, this empirical parameterization is only valid for
 intermediate values of $x$; we fit only including data in the range $0.1 \leq x \leq 0.7$.  In this
 range, this is clearly a good description of the lattice data.  Numerical results for the fit
 parameters $A_H$ and $B_H$ for various quantities are presented in Table~\ref{tab:x_fits}.

For the heaviest quark masses in Fig.~\ref{fig:meson}, significant curvature is evident, particularly
 when including the physical charmonium states.  This behavior is to be expected; the horizontal axis
 $(M_{PS}/M_V)^2$ goes to 1 in the limit $m_q \rightarrow \infty$, but in the same limit the hadron mass
 on the vertical axis will also go to infinity.  Indeed, at heavy quark mass we expect the contribution 
to hadron masses to be dominated by the quark masses themselves, so that we should expect linear
 behavior in $m_q$ instead.  This is clearly shown in the bottom panel of the figure, and we include
 the results of a simple linear fit
\bee
M_H = C_H + D_H m_q
\ee
where $m_q$ is the quark mass in MeV.  We restrict the fits to include lattice data 
with $m_q > 200$ MeV; only our own lattice simulation results are included in this region.  
This has the benefit of giving a consistent treatment of quark-mass renormalization: our $m_q$ are
 perturbatively renormalized in the $\overline{MS}$ scheme at a  scale $\mu=2$ GeV \cite{DeGrand:2002vu}. 
 Numerical results for $C_H$ and $D_H$ are collected for various quantities in Table~\ref{tab:mq_fits}.

\begin{figure}
\begin{center}
\includegraphics[width=0.7\textwidth,clip]{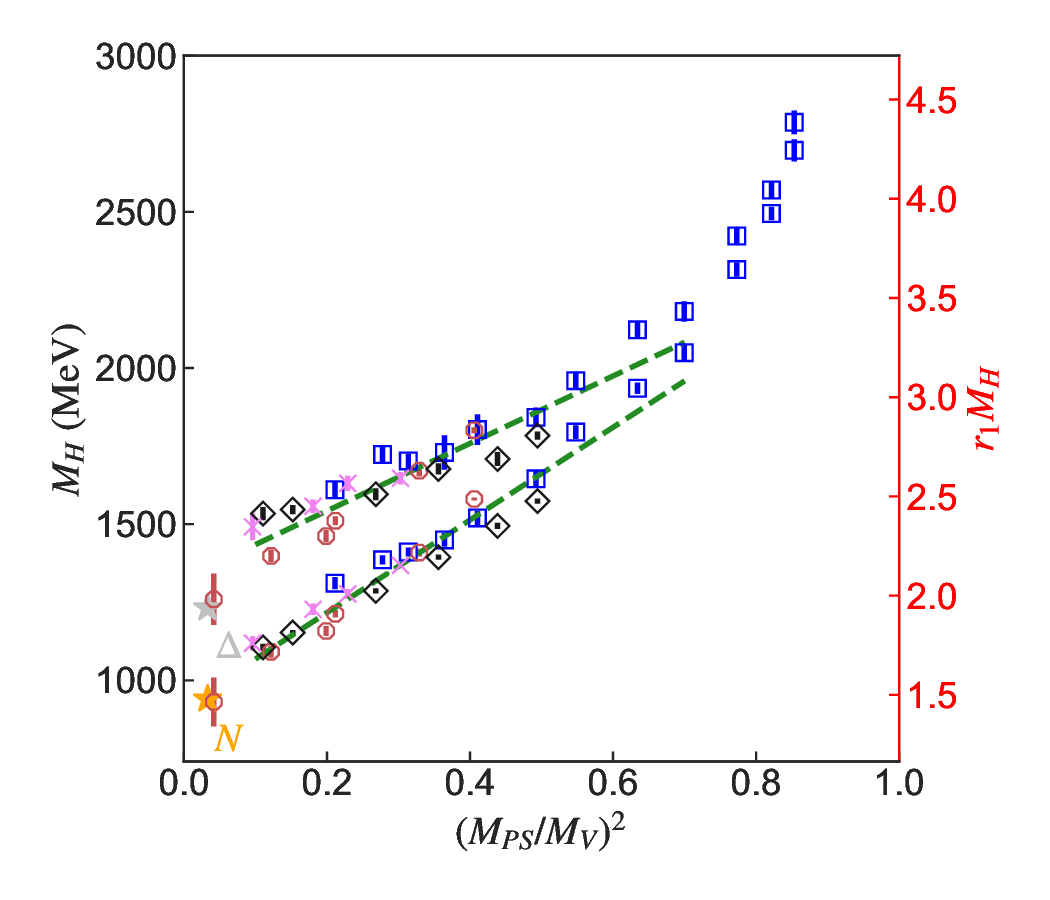}
\includegraphics[width=0.7\textwidth,clip]{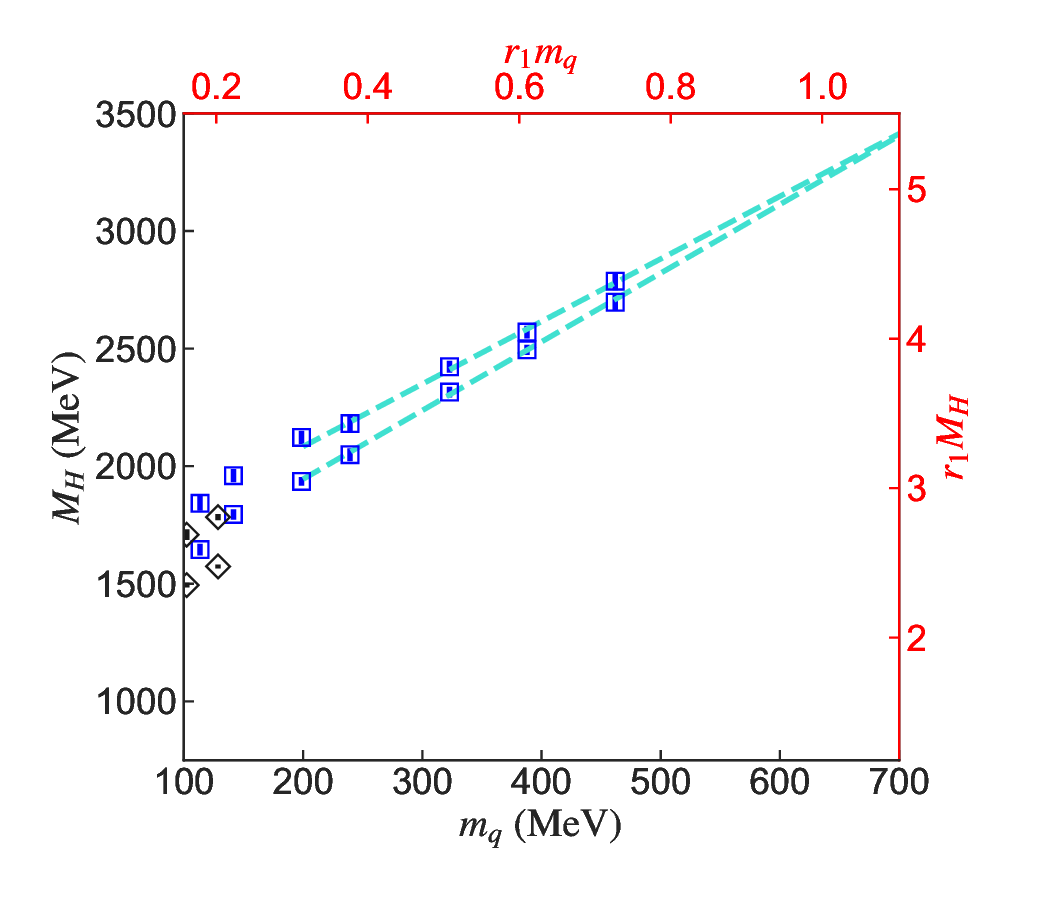}
\end{center}
\caption{Nucleon (lower band) and delta baryon 
(upper band) masses in MeV, as a function of $(M_{PS}/M_V)^2$ (top panel) and quark mass
in MeV (bottom panel).
Data are black diamonds from Ref.~{\protect{\cite{WalkerLoud:2008bp}}},
red octagons from Ref.~{\protect{\cite{Aoki:2008sm}}},
violet crosses from Refs.~{\protect{\cite{Alexandrou:2008tn,Jansen:2009hr,Baron:2009wt}}},
blue squares from this work.  Stars show the physical nucleon and delta masses, and dashed lines show
linear fits to the data as described in the text.
\label{fig:baryon}}
\end{figure}

Next, we turn to the nucleon and delta states, shown in Fig.~\ref{fig:baryon}.
We have added the physical values of the masses of the nucleon and delta to the plot.
As in the meson case, we have included empirical fits as a function of $x$ (for intermediate quark masses)
and as a function of $m_q$ (for heavy quark masses).

\begin{figure}
\begin{center}
\includegraphics[width=0.8\textwidth,clip]{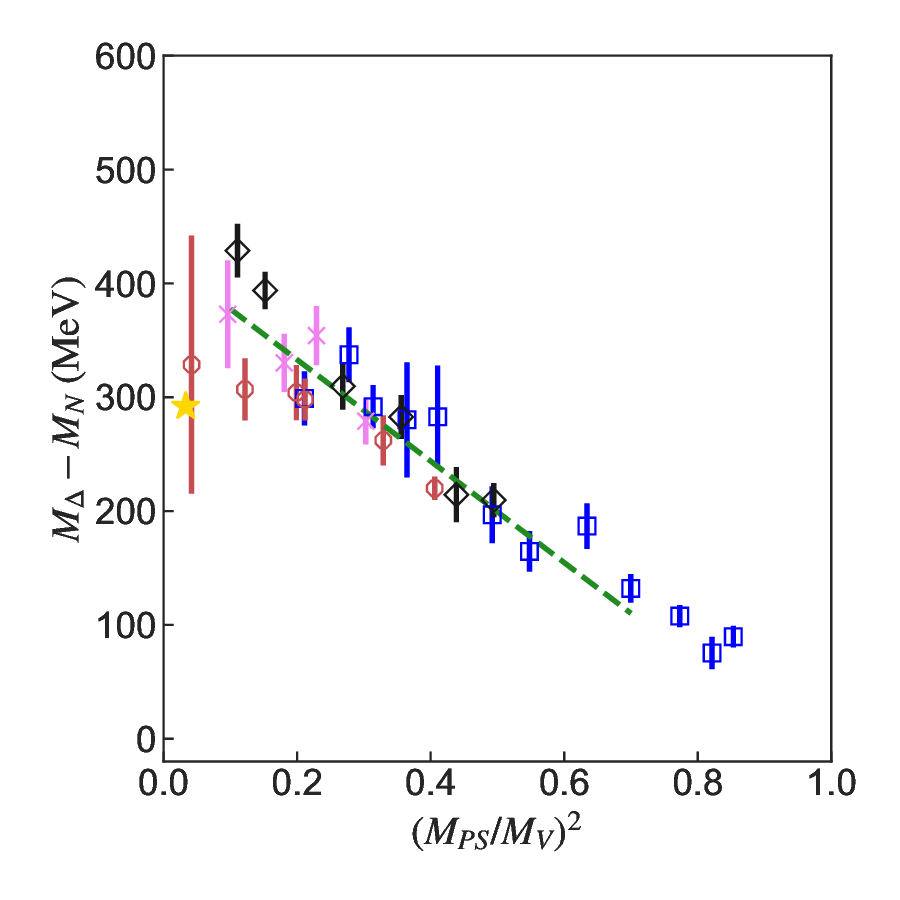}
\end{center}
\caption{Delta-nucleon mass difference in MeV as a function of $(M_{PS}/M_V)^2$.
Data are black diamonds from Ref.~{\protect{\cite{WalkerLoud:2008bp}}},
red octagons from Ref.~{\protect{\cite{Aoki:2008sm}}},
violet crosses from Refs.~{\protect{\cite{Alexandrou:2008tn,Jansen:2009hr,Baron:2009wt}}},
blue squares from this work.  The star shows the physical point, and the dashed line shows a linear
fit to the data.
\label{fig:dn_diff}}
\end{figure}

 We can obtain a
clearer picture by considering the mass difference $M_\Delta - M_N$ directly; this quantity 
is shown in Fig.~\ref{fig:dn_diff}.
In quark models, we expect that the delta-nucleon mass splitting should vanish as $m_q \rightarrow \infty$.
How it vanishes is model dependent. In models where the splitting is given by
a color hyperfine splitting, it would go as a product of the two color magnetic moments (and thus would scale as 
$1/m_q^2$ times a wave function factor). Our lattice data are not precise enough, nor do
they extent to large enough quark masses, to say more about this point.

We can extract additional information from the slope of the nucleon mass with respect to the
 quark mass.  From the Feynman-Hellmann theorem, the derivative $\partial M_N / \partial m_q$ yields
 the scalar matrix element $\bra{N}  \bar{q} q \ket{N}$.  A more practically useful definition of this quantity is in
 terms of the baryon ``sigma term'',
\bee
f_S^{(N)} \equiv \frac{\bra{N} m_q \bar{q} q \ket{N}}{M_N} = \frac{m_q}{M_N} \frac{\partial M_N}{\partial m_q},
\ee
which cancels out the quark-mass renormalization dependence.  The sigma term is of particular 
interest in describing interactions of the Higgs boson with the baryon, either in QCD or in beyond 
standard model scenarios.  For example, it may be used to constrain the Higgs portal interaction of
baryon-like dark matter candidates \cite{Appelquist:2014jch}.

\begin{figure}
\begin{center}
\includegraphics[width=0.8\textwidth,clip]{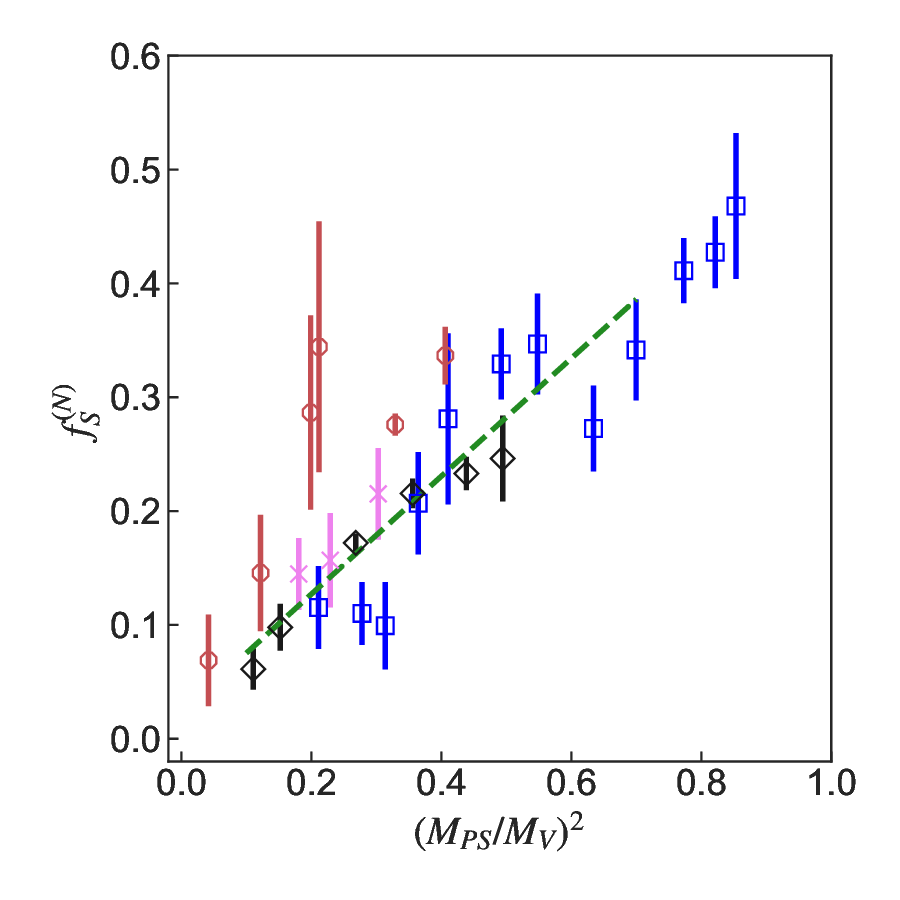}
\end{center}
\caption{Scalar form factor of the nucleon as a function of $(M_{PS}/M_V)^2$.
Data are black diamonds from Ref.~{\protect{\cite{WalkerLoud:2008bp}}},
red octagons from Ref.~{\protect{\cite{Aoki:2008sm}}},
violet crosses from Refs.~{\protect{\cite{Alexandrou:2008tn,Jansen:2009hr,Baron:2009wt}}},
blue squares from this work.  The fit shown (dashed line) includes a 10\% systematic error on all points, in order to avoid bias due to outliers with very small error bars.
\label{fig:sigma}}
\end{figure}

We determine $f_S^{(N)}$ directly from the lattice data using a second-order finite difference 
approximation to the derivative; our results are shown in Fig.~\ref{fig:sigma}.  Although there
 are some outlier points with anomalously small errors, for the most part we see universal 
agreement of the lattice results with linear behavior in the regime $0.1 \leq (M_{PS} / M_V)^2 \leq 0.7$. 
 The curve obtained here is similar to the results seen at larger $N_c$ and even with different
 color-group representations for the quarks, as discussed in \cite{DeGrand:2015lna,Kribs:2016cew}.

\begin{table}
\begin{tabular}{c c c }
\hline
observable & $A_H$ & $B_H$ \\
\hline
$M_V$ (MeV) & 760 & 720 \\
$M_A$ (MeV)& 1120 & 1040 \\
$M_N$ (MeV) & 920 & 1480 \\
$M_\Delta$ (MeV) & 1330 & 1080 \\
$M_\Delta - M_N$ (MeV) & 422 & -446 \\ 
$f_S^{(N)}$ & 0.02 & 0.52 \\
$f_{PS}$ (MeV) & 134 & 117 \\
$M_{PS} / f_{PS}$ & 1.61 & 4.86 \\
$f_V$ & 0.299 & -0.081 \\
$f_A$ & 0.218 & -0.100 \\
\hline
 \end{tabular}
\caption{ Simple parameterization of hadronic observables, using a linear model of the form $A_H + B_H x$ in $x \equiv (M_{PS}/M_V)^2$.  All dimensionful quantities are given in units of MeV.  As discussed in the text, this parameterization should only be used in the range $0.1 \leq x \leq 0.7$.  No error bars are given, but this parameterization should be accurate at roughly the 10\% level, as may be seen from the plots. \label{tab:x_fits}}
\end{table}

\begin{table}
\begin{tabular}{c c c }
\hline
observable & $C_H$ & $D_H$ \\
\hline
$M_V$ (MeV) & 960 & 1.71 \\
$M_A$ (MeV)& 1450 & 1.65 \\
$M_N$ (MeV) & 1360 & 2.92 \\
$M_\Delta$ (MeV) & 1550 & 2.66 \\
$M_{PS} / f_{PS}$ & 3.87 & $3.99 \times 10^{-3}$ \\
$f_V$ & 0.266 & -0.131 \\
$f_A$ & 0.179 & -0.142 \\
\hline
 \end{tabular}
\caption{ Alternative parameterization of hadronic observables, using a linear model of the form 
$C_H + D_H m_q$.  All dimensionful quantities (including $m_q$) are given in units of MeV.  As
 discussed in the text, this parameterization should only be used for $m_q > 200$ MeV.  No error bars 
are given, but this parameterization should be accurate at roughly the 10\% level, as may be seen from the plots. \label{tab:mq_fits}}
\end{table}

\subsection{Other states \label{sec:other}}

Here we briefly discuss and review available lattice QCD results for other states, including 
excited states, higher-spin states, glueballs, and other exotica.

 \subsubsection{Excited states and higher spin}

Lattice data for excited states and states with higher spin is much sparser than for ground state hadrons. We
 can point the reader at Ref.~\cite{Dudek:2013yja}
for a calculation of a variety of such states at $(M_{PS}/M_V)^2=0.43, 0.29$ and 0.19, and
 Ref.~\cite{Engel:2011aa} for a range of mass values, roughly $0.1 \lesssim (M_{PS}/M_V)^2 \lesssim 0.3$.

\subsubsection{Glueballs}

In some phenomenological scenarios, such as the Fraternal Twin Higgs model of Ref.~\cite{Cheng:2015buv},
 the quarks are truly heavy, so that the spectrum of such models consists entirely of light glueballs
 and quarkonia.  The heavy-quark states in this case may be identified as ``quirks'' \cite{Kang:2008ea}, exhibiting the formation of macroscopic color-force strings; we will not explore this scenario in detail here, but we direct the interested reader to \cite{Lucini:2004my,Teper:2009uf} for lattice results on string formation in pure gauge theory in the large-$N_c$ limit.
 
 In the heavy-quark limit, the glueball spectrum is basically that of pure-gauge (``quenched'') QCD,  rescaled appropriately. Lattice data is available for this spectroscopy: a definitive study for SU(3) is e.~g.~Ref.~\cite{Morningstar:1999rf}), while
 for a study of the large-$N_c$ limit we direct the reader to Refs.~\cite{Lucini:2010hn,Lucini:2010nv}.  
See also the review Ref.~\cite{McNeile:2008sr}.

Including the effects of quark masses, the study of glueballs becomes more difficult, due to severe signal-to-noise problems that require the high statistics most easily obtained in pure-gauge calculations.  Ref.~\cite{Gregory:2012hu} presents glueball spectrum results with a pion mass of 360 MeV (or $(M_{PS} / M_V)^2 \sim 0.17$), which may give some sense of how the glueball spectrum changes away from the pure-gauge limit.  We will not attempt to review the lattice QCD literature on attempting to calculate glueball masses at the physical point, which is certainly an open research question.

\subsubsection{Exotic states}

Finally, there are true exotic states, involving the QCD properties of particles that do not exist in the standard model,
 for example fermions charged under higher representations of SU(3) than the fundamental.  These are somewhat beyond
 the scope of our study, as such states are not part of ordinary lattice QCD calculations and so there is not a
 wealth of results available to repurpose.  However, these exotic states can be important in certain BSM scenarios,
 so we will briefly review some available results. 

Fermions charged under the adjoint representation of SU(3) are natural to consider; they appear as gluinos in 
supersymmetric extensions of the standard model, or in other scenarios such as the ``gluequarks'' of Ref.~\cite{Contino:2018crt}.  In cases where the adjoint fermion is stable, it can form QCD bound states whose binding can be studied on the lattice.  An early important work studying adjoint fermions is
 Ref.~\cite{Foster:1998wu}, working in the quenched approximation (i.~e.~ effectively at infinite
 quark mass.)  A more recent study is Ref.~\cite{Marsh:2013xsa}, which also considers fermions in 
representations as high as the $\textbf{35}$ of SU(3), working at $(M_{PS} / M_V)^2 \sim 0.38$.

There is a vast literature on the inclusion of \emph{light} fermions in higher representations, for the
 purposes of studying the transition to infrared-conformal behavior: see \cite{DeGrand:2015zxa, Svetitsky:2017xqk}
 for recent reviews of this subfield.  These systems are qualitatively different from QCD and so we will not
 say anything further about them here, except to note that the possibility of vastly different infrared
 behavior, or even loss of asymptotic freedom, should not be forgotten by model-builders including 
fermions in higher representations.

Another exotic possibility would be heavy fundamental (i.e.\ not composite) scalars which carry SU(3) color charge in some representation - again, supersymmetric theories give rise to squarks as a prototypical example.  This could  lead to a wealth of interesting scalar-quark bound states (e.g.\ ``R-hadrons'' \cite{Farrar:1978xj,Kraan:2004tz}.)  We can find very little on this subject in the  literature, although certainly there is a significant amount of early work on the inclusion of light
 scalars and phase diagrams of scalar-gauge theories.  Relevant to the current context we can find only
 Ref.~\cite{Iida:2007qp}, which studies the spectrum of bound states including scalars in the fundamental
 representation of SU(3); they use the quenched approximation and extremely coarse lattice spacing with
 no continuum extrapolation, so their results should be applied with appropriate caution.

\section{Vacuum transition matrix elements \label{sec:matel} }

One of the simplest matrix elements to compute on the lattice is the matrix element for decay of a hadronic
 state to the vacuum state through an intermediate operator,  $\bra{0} \mathcal{O} \ket{H}$.  The presence of
 only a single strongly-coupled state means that these quantities can be calculated with simple correlation
 functions and good signal-to-noise compared to processes involving multiple hadronic states.

\subsection{Pseudoscalar decay constant}
The \emph{decay constants} parameterize specific vacuum transition matrix elements of the pseudoscalar,
 vector, and axial vector mesons.  They have an extensive history in lattice simulations; we begin with
 the pseudoscalar meson.  Introducing the quark flavor labels $u,d$
to characterize the current, the pseudoscalar decay constant $f_{PS}$ is defined through
\bee
\langle 0| \bar u \gamma_0 \gamma_5 d |\pi\rangle = M_{PS} f_{PS}.
\ee
Our conventions lead to the identification $f_{\pi} \sim 130$ MeV in QCD, but it should be emphasized 
that this choice is not unique.  Although the decay width of the pion is a physical and experimentally 
accessible quantity, the decay constant $f_{PS}$ is not physical in the same sense; its precise value depends
 on various choices of convention; a detailed discussion is given in Appendix~\ref{sec:fPS}.   We show our summary of lattice results for $f_{PS}$ in Fig.~\ref{fig:fpi}.

\begin{figure}
\begin{center}
\includegraphics[width=0.8\textwidth,clip]{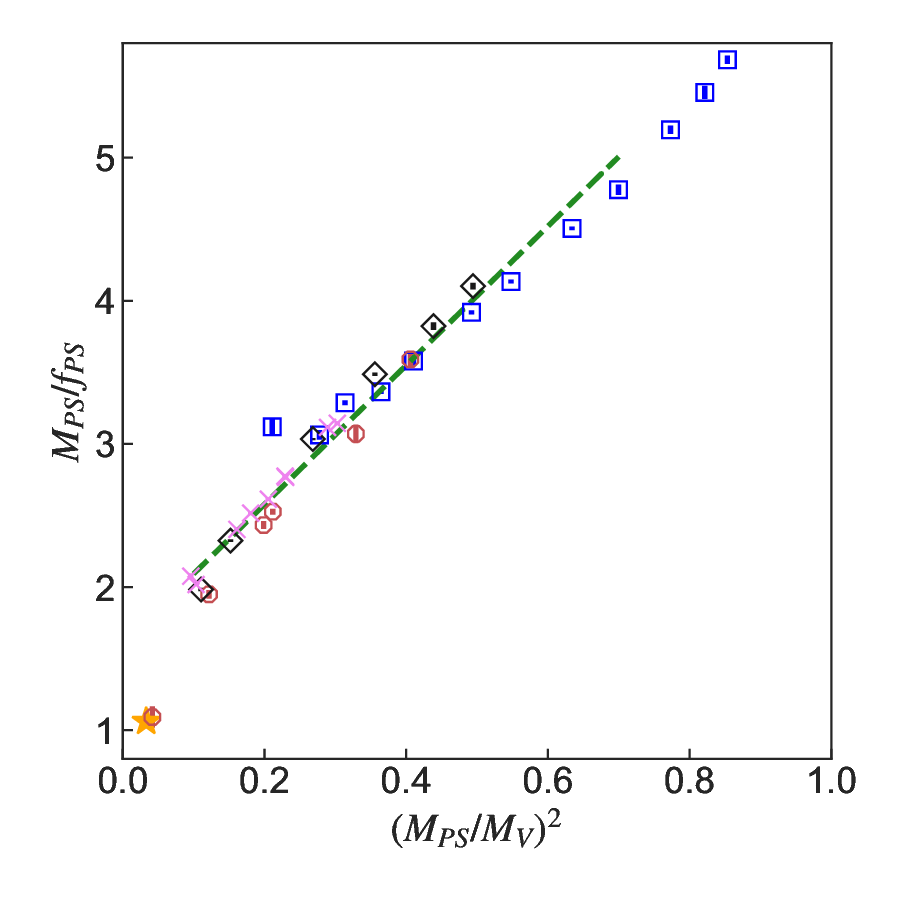}
\end{center}
\caption{Ratio of pseudoscalar mass to decay constant  as a function of $(M_{PS}/M_V)^2$.
Data are black diamonds from Ref.~{\protect{\cite{WalkerLoud:2008bp}}},
red octagons from Ref.~{\protect{\cite{Aoki:2008sm}}},
violet crosses from Refs.~{\protect{\cite{Alexandrou:2008tn,Jansen:2009hr,Baron:2009wt}}},
blue squares from this work.
\label{fig:mpifpi}}
\end{figure}

\begin{figure}
\begin{center}
\includegraphics[width=0.8\textwidth,clip]{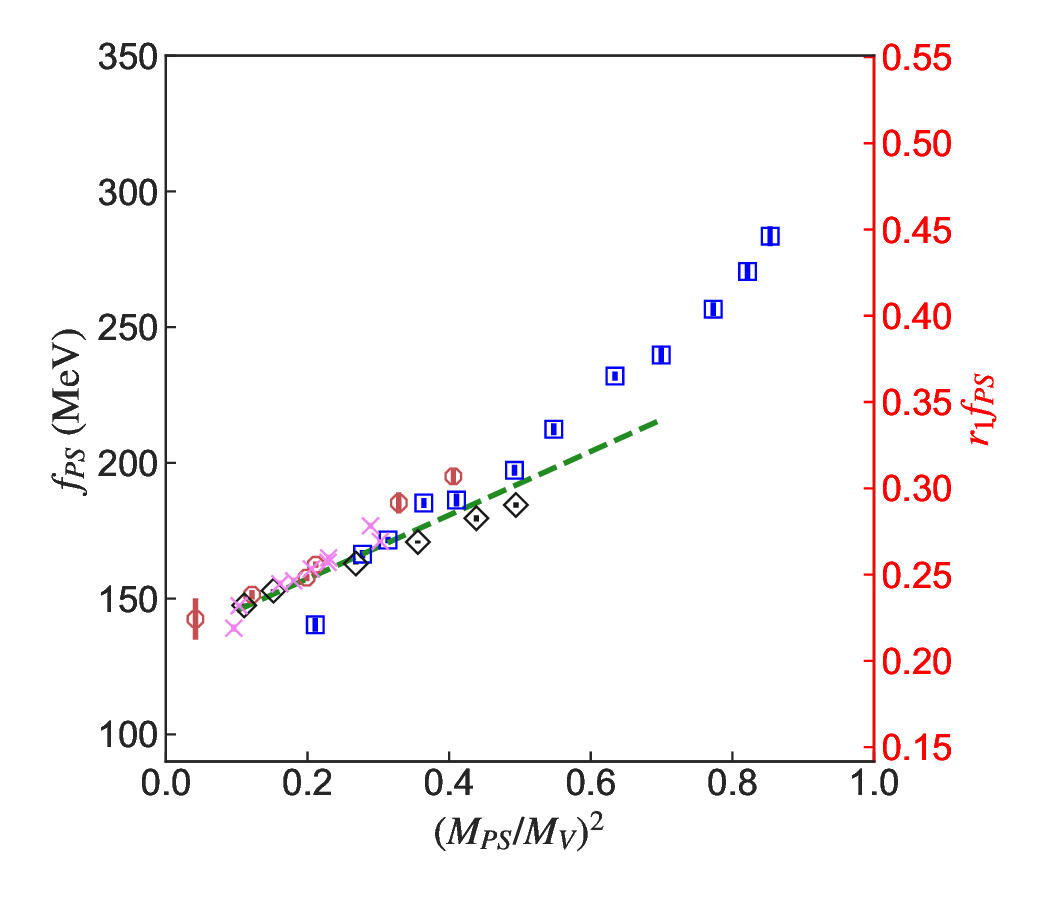}
\end{center}
\caption{Pseudoscalar decay constant  as a function of $(M_{PS}/M_V)^2$..
Data are black diamonds from Ref.~{\protect{\cite{WalkerLoud:2008bp}}},
red octagons from Ref.~{\protect{\cite{Aoki:2008sm}}},
violet crosses from Refs.~{\protect{\cite{Alexandrou:2008tn,Jansen:2009hr,Baron:2009wt}}},
blue squares from this work.
\label{fig:fpi}}
\end{figure}

Perhaps a more useful quantity is the ratio $M_{PS}/f_{PS}$. It sometimes appears as a free parameter
in the phenomenological literature, where it is allowed to vary over a large
range. 
(For example, see Fig.~1 of Ref.~\cite{Hochberg:2014kqa}.)
 We show this ratio in Fig.~\ref{fig:mpifpi}.
 In QCD, the range of possible values for this ratio is quite limited, ranging from about 1 at
 the physical point to 5-6 for the heaviest quark masses we consider. 
 
 In heavy quark effective theory, there is a solid expectation that $f_{PS}$ is proportional
to $1/\sqrt{M_{PS}}$ in the asymmetric limit that one quark becomes very heavy.
 This is confirmed by lattice simulations.
For two degenerate
masses,  quark models suggest that this result is modified to $f_{PS} \sim 1/\sqrt{M_{PS}} \psi(0)$
where $\psi(0)$ is the meson wave function at zero separation.  We are unaware of direct lattice checks
of a degenerate-mass decay constant at very large quark masses.

\subsection{Vector and axial-vector decay constants}

The vector meson decay constant of state $V$ is defined as
\bee
\langle 0| \bar u \gamma_i d  | V\rangle = M_V^2 f_V \epsilon_i
\ee
and the axial vector meson decay constant of state $A$ is  defined as
\bee
\langle 0|  \bar u \gamma_i \gamma_5 d  |A \rangle = M_A^2 f_A \epsilon_i.
\ee
where $\epsilon_i$ is a unit polarization vector.  Once again, we emphasize that conventions for the 
definition of these decay constants vary in the literature; in particular, with our definitions
 $f_V$ and $f_A$ are dimensionless, but dimensionful versions of the decay constants are commonly used as well.  These quantities appear in the phenomenological literature both in the coupling of bound states
to photons and $W$'s, and also in their coupling to new gauge bosons such as dark photons, although the decay constants often do not appear directly.

We could not find a full set of axial vector meson decay constants in the literature, and so we generated our own
data. The signal in the axial vector channel is much noisier than in the vector channel, and so we used
data sets of 400 stored configurations, run at some of the same simulation parameters as was used in Ref.~\cite{DeGrand:2016pur}.
The analysis is identical to the one done in that paper.

\begin{figure}
\begin{center}
\includegraphics[width=0.8\textwidth,clip]{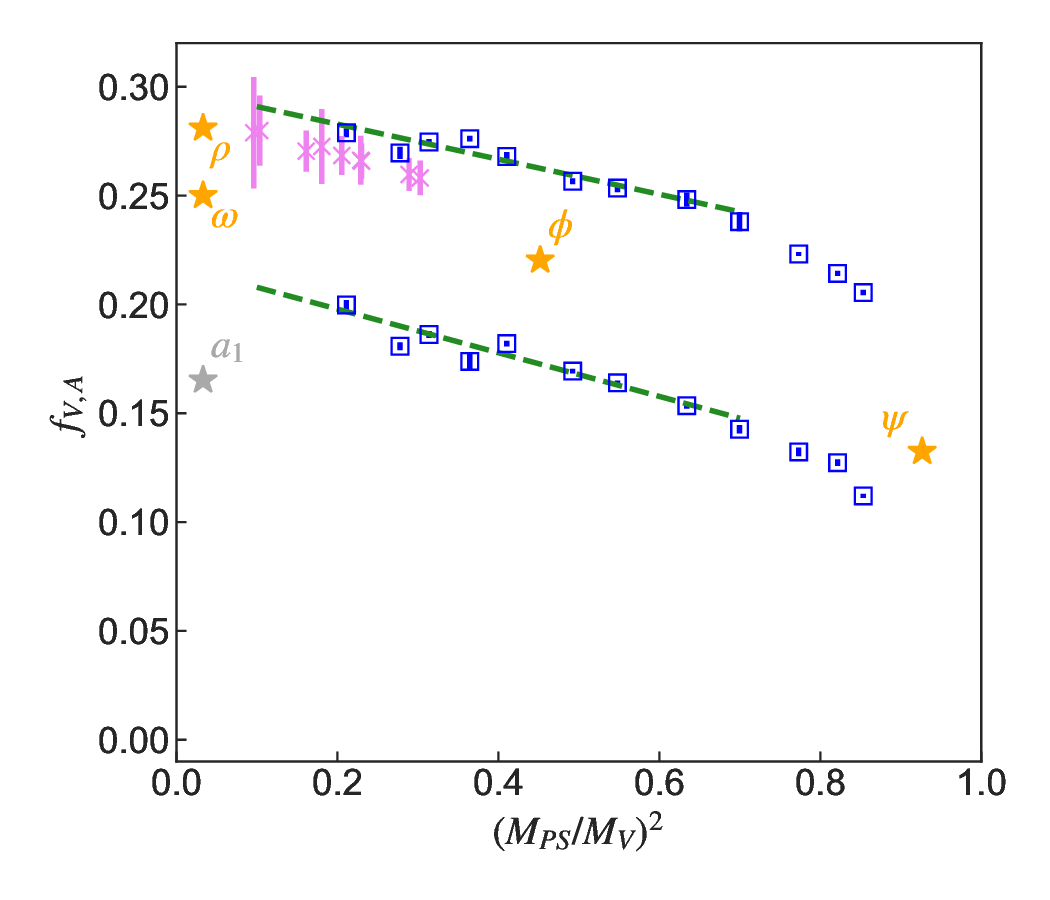}
\end{center}
\caption{Vector meson (upper band) and axial vector meson (lower band) decay constants versus
$(M_{PS}/M_V)^2$. Data are violet crosses from Ref.~{\protect{\cite{Jansen:2009hr}}} and blue squares from this work.
  Stars show physical values for various states, determined as described in the text.\label{fig:decaynewpirho}}
\end{figure}

We show the decay constants as a function of $(M_{PS}/M_V)^2$ in Fig.~\ref{fig:decaynewpirho}.
We overlay decay constants inferred from experimental data from the 
Review of Particle Properties \cite{Patrignani:2016xqp}.
In our conventions, the vector meson decay width to electrons is
\bee
\Gamma(V \rightarrow e^+ e^-) = \frac{4\pi\alpha^2}{3} M_V f_V^2 \svev{q}^2
\ee
where $\svev{q}$ is the average charge: $-1/3$ for the phi meson, $(2/3- (-1/3))/\sqrt{2}$ for the rho,
 and $(2/3 +(-1/3))/\sqrt{2}$ for the omega.  Extracting a decay constant from the $a_1$ is complicated by 
its large width, so we take the phenomenological result from the old analysis of Ref.~\cite{Isgur:1988vm}. 
 Significant deviations of these experimental values from the lattice data are seen, on the order of 20\%;
 this may reflect large systematic uncertainties in our determinations of $f_V$ and $f_A$ which are not accounted for.

As a concrete example of applying vector decay constants, we consider the composite dark sector model of Ref.~\cite{Harigaya:2016rwr}.  In section V, the reference discusses the generation of a coupling $\epsilon'$ of the ``dark rho'' meson to the standard model electromagnetic current, $\mathcal{L} = \epsilon' \rho_{D,\mu} J_{\rm em}^\mu$.  This interaction is described as arising from mixing of the dark rho with a dark photon $A_D^\mu$, which in turn mixes with the ordinary photon with mixing strength $\epsilon$.  Carrying out a simple effective matching, the mixing of dark rho with dark photon is given by $\langle A_D | \rho_D \rangle = e_D \langle 0 | j_V^{\mu} | \rho_D \rangle = e_D M_{\rho_D}^2 f_{\rho_D}$.  Since $M_{\rho_D} \gg M_{A_D}$ in this model, the dark photon propagator cancels the $M_{\rho_D}^2$, leaving the result
\bee
\epsilon' = \epsilon e_D f_{\rho_D} \approx \epsilon e_D \frac{\sqrt{N}}{4\sqrt{3}},
\ee
where $N$ is the number of colors in the SU$(N)$ dark sector.  Here we have substituted the result $f_V \approx 1/4$ from our collected lattice results and made use of the known large-$N$ scaling of decay constants as $\sqrt{N}$ \cite{Bali:2013kia}.  Compared to the naive dimensional analysis (NDA) result given in the reference, this estimate is larger by a factor of $\pi / \sqrt{3}$ - nearly a factor of two.  A similar analysis can be applied to the NDA estimates of dark rho couplings in \cite{Kribs:2018ilo}.

\subsection{Sum-rule relations between vector and pseudoscalar properties}

The properties of the vector ($\rho$) meson can be closely related in QCD to certain interactions involving pions,
 particularly the pion electromagnetic form factor (this idea is known as vector meson dominance, or VMD; see
 Ref.~\cite{Klingl:1996by} for a good review.)  The modern approach to this idea is couched in extensions of
 the chiral Lagrangian, but early work on relating vector and pseudoscalar interactions was done in the framework 
of current algebra, one of the highlights of which are the KSRF relations (after Kawarabayashi, Suzuki, Riazuddin, and
Fayazuddin \cite{Kawarabayashi:1966kd,Riazuddin:1966sw}.)

For the vector meson decay constant, KSRF predicts (in our conventions) that
\bee 
f_V = \sqrt{2} \frac{f_{PS}}{M_V}.
\label{eq:ksrfv}
\ee

\begin{figure}
\begin{center}
\includegraphics[width=0.8\textwidth,clip]{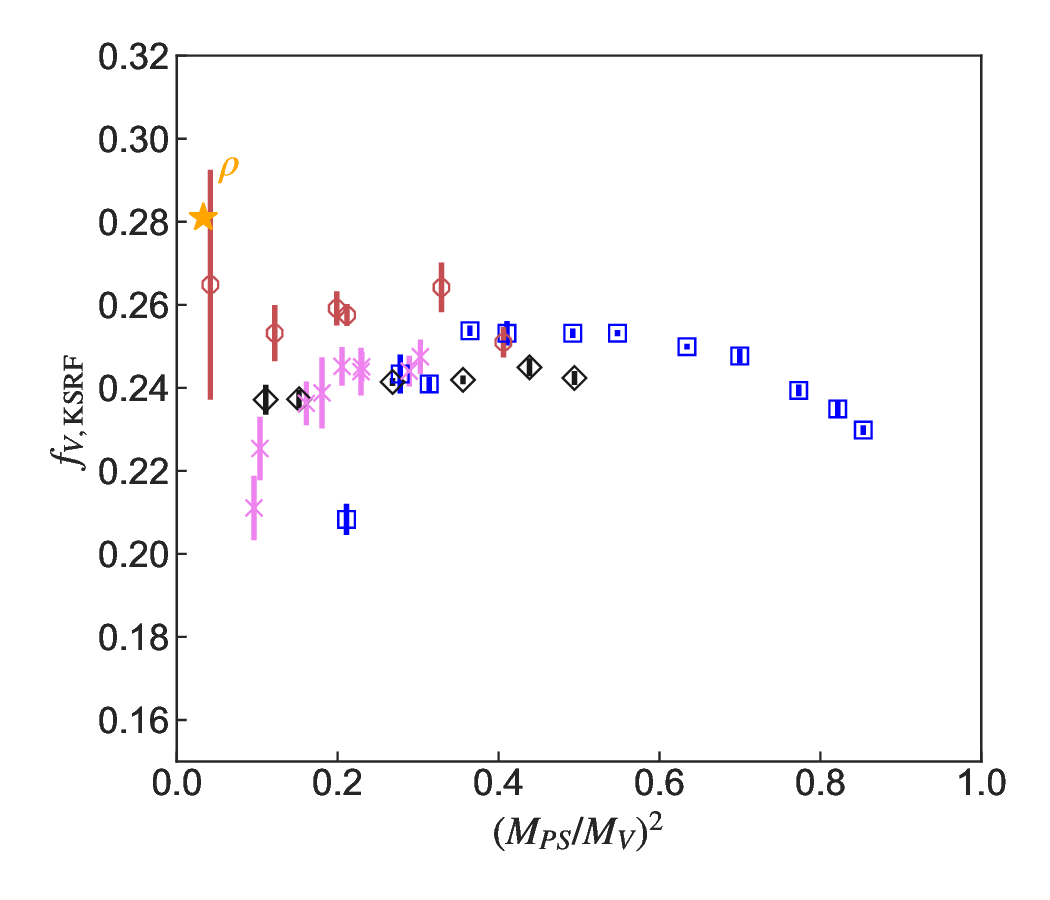}
\end{center}
\caption{
Vector meson decay constant
$f_V$ versus
$(M_{PS}/M_V)^2$ as inferred from the KSRF relations.
Data are black diamonds from Ref.~{\protect{\cite{WalkerLoud:2008bp}}},
red octagons from Ref.~{\protect{\cite{Aoki:2008sm}}},
violet crosses from Refs.~{\protect{\cite{Alexandrou:2008tn,Jansen:2009hr,Baron:2009wt}}},
blue squares from this work.  The star shows the physical $\rho$ meson decay constant.
\label{fig:ksrfv}}
\end{figure}

A comparison of this relation is shown in  Fig.~\ref{fig:ksrfv}.
The parameterization seems to work qualitatively well at intermediate masses; there is some tension between the KSRF and direct results at light masses, but the disagreement between different lattice groups here indicates the possible onset of systematic effects.  In particular, we note that the value of $f_V$ inferred from this ratio changes very little with $(M_{PS} / M_V)^2$, indicating weak dependence on the quarks; similar results have been seen in other theories with additional light quarks, see for example Refs.~\cite{Appelquist:2016viq,Nogradi:2019iek}.

In a theory with spontaneous chiral symmetry breaking, the pseudoscalar decay constant
and sums of the vector and axial vector decay constants are constrained by the
first
\bee
\sum_V f_V^2M_V^2 - \sum_A f_A^2M_A^2 -f_{PS}^2 =0
\label{eq:W1}
\ee
and second
\bee
\sum_V f_V^2M_V^4 - \sum_A f_A^2 M_A^4 =0
\label{eq:W2}
\ee
 Weinberg sum rules \cite{Weinberg:1967kj}. One often sees phenomenology where
the difference of vector and axial spectral functions is saturated by the three lowest
states (the pion, rho and $a_1$) and the decay constants are constrained to satisfy the
 Weinberg sum rules (typically by fixing $f_{a_1}$. This is called the ``minimal hadron approximation''
\cite{Peskin:1980gc}.

\begin{figure}
\begin{center}
\includegraphics[width=0.8\textwidth,clip]{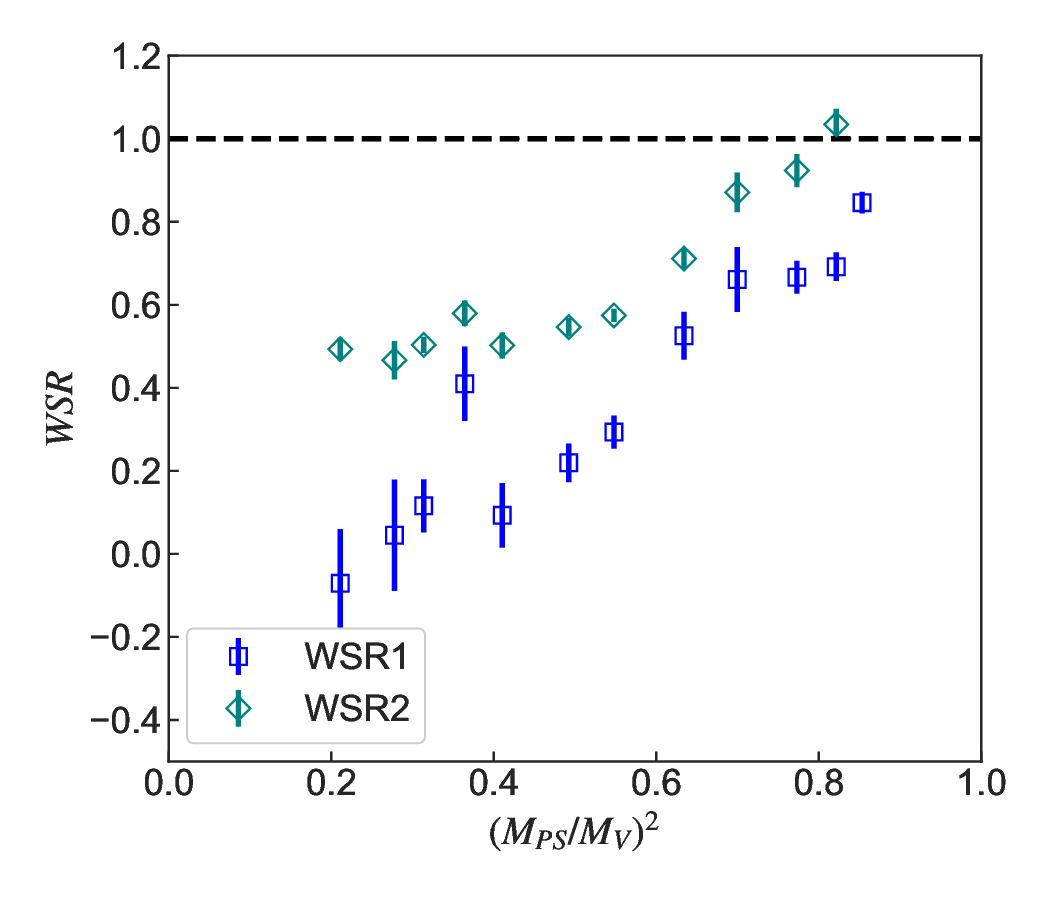}
\end{center}
\caption{Tests of the Weinberg sum rules, equations~(\ref{eq:W1}) and (\ref{eq:W2}), using the lattice data from this work.  Both quantities should be equal to 1 if the Weinberg sum rules hold.  Significant deviations are seen at essentially all values of the quark mass for both sum rules.
\label{fig:wsr}}
\end{figure}

 Such an approximation is not justified by 
the lattice-determined decay constants.  In Fig.~\ref{fig:wsr}, we present two quantities which test the 
Weinberg sum rules using the lowest states: ``WSR1'' denotes the combination
 $(f_V^2 M_V^2 - f_A^2 M_A^2) / f_{PS}^2$, while ``WSR2'' is the expression $(f_V^2 M_V^4) / (f_A^2 M_A^4)$.  As can
 be seen from the plot, significant deviations from the expected result of 1 are seen over a wide range
 of quark masses.  As a result, we caution against the use of the Weinberg sum rules with minimal hadron approximation.

\subsection{Nucleon decay matrix elements}

There is a smaller literature on vacuum-transition matrix elements for the nucleons.  For the study of
 proton decay in QCD, the more interesting decay processes include a pion in the final state,
 e.~g.~$p \rightarrow \pi^0 e^+$.  However, very early lattice work attempted to estimate this more 
complicated matrix element in terms of the simpler proton-to-vacuum matrix elements (this was known as the ``indirect method''.)

Instead of attempting to review the literature, we refer to the recent lattice result Ref.~\cite{Aoki:2017puj}.
  They define the low-energy constants $\alpha$ and $\beta$ as
\begin{align}
\bra{0} (ud)_R u_L \ket{p^+} &= \alpha P_R u_p, \\
\bra{0} (ud)_L u_L \ket{p^+} &= \beta P_L u_p,
\end{align}
which are precisely the vacuum matrix elements.  Their study uses a single lattice spacing and four ensembles
 with $340\ {\rm MeV} \lesssim M_{PS} \lesssim 690\ {\rm MeV}$, corresponding roughly to 
$ 0.15 \lesssim (M_{PS} / M_V)^2 \lesssim 0.40$.  They find
\bee
\alpha \approx -\beta = -0.0144(3)(21)\ {\rm GeV}^3
\ee
in $\overline{MS}$ renormalized at $\mu = 2$ GeV, extrapolated to the physical point.  The data for mass dependence 
of this quantity is not presented directly, but their figure 2 shows that the unrenormalized results for 
$\alpha$ and $\beta$ at the heavier quark masses are larger by up to a factor of 2.

\section{Strong decays \label{sec:decay}}
Of course, all QCD states which can decay strongly will do so. This physics can be important for phenomenology.
An example is the decay of dark matter particles, described in Ref.~\cite{Berlin:2018tvf}.  Lattice calculations extract coupling constants for decay processes indirectly. The calculation begins 
by finding the masses of multistate systems in a finite box; the shift in mass parameterizes the
interaction between the particles.
Ref.~\cite{Briceno:2017max} is a good recent review.  We will briefly consider results for the vector
 and scalar mesons as $\pi \pi$ resonances; this field is relatively new within lattice QCD, and so
 there are very few results yet for resonances associated with other combinations of mesons.

\subsection{$\rho\rightarrow \pi\pi$}
The most extensive lattice results are for the rho meson.
In contrast, phenomenology often uses a KSRF relation for
 the coupling constant mediating the decay of a vector into two pseudoscalars,
\bee
g_{VPP} = \frac{M_V}{f_{PS}}.
\ee
The vector meson decay width is
\bee 
\Gamma(V \rightarrow PP) \simeq \frac{g^2_{VPP} }{48 \pi M_V^2}(M_V^2-4M_{PS}^2)^{3/2} .
\label{eq:vector_width}
\ee

\begin{figure}
\begin{center}
\includegraphics[width=0.8\textwidth,clip]{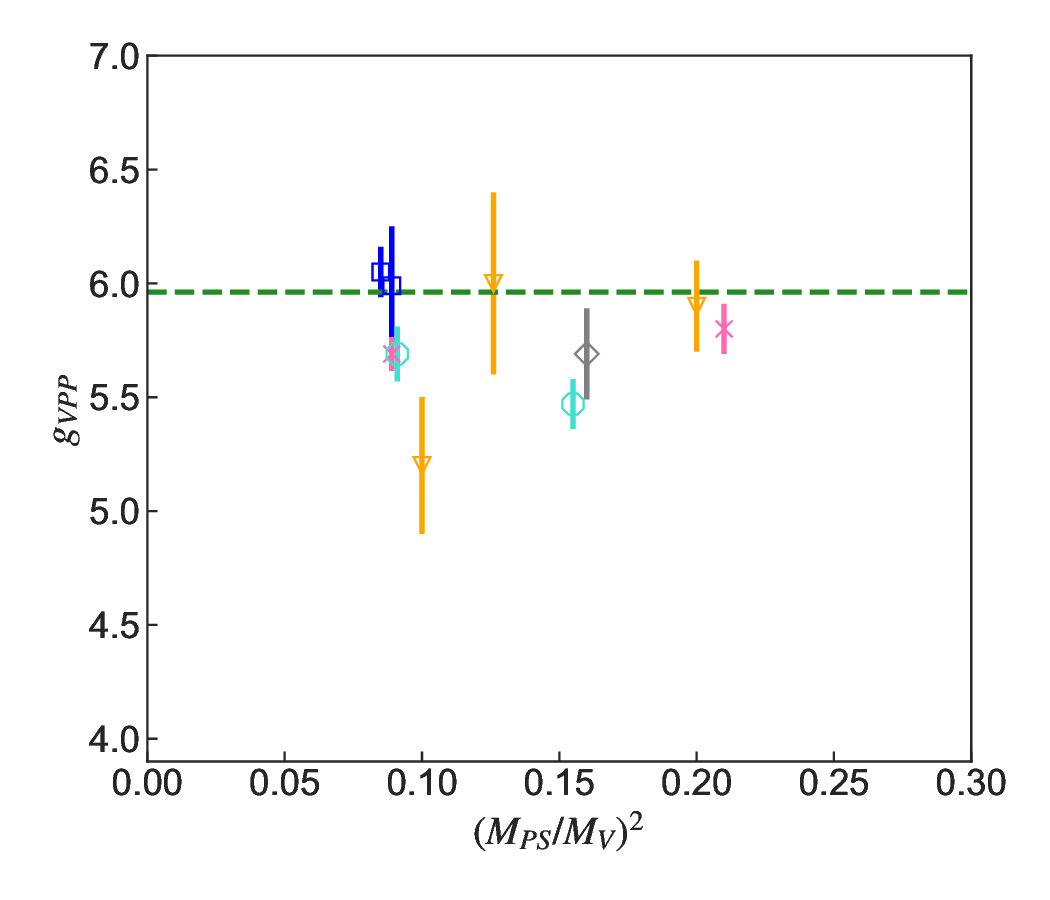}
\end{center}
\caption{The vector meson decay constant $g_{VPP}$ from lattice calculations,
 as a function of $(M_{PS}/M_V)^2$.
 Symbols are
blue squares, Ref.~\cite{Bulava:2016mks} and \cite{Bulava:2017stw};
pink crosses, Ref.~\cite{Dudek:2012xn} and \cite{Wilson:2015dqa};
light blue octagons, Ref.~\cite{Guo:2016zos};
grey diamond, Ref.~\cite{Alexandrou:2017mpi}
and yellow triangles, Ref.~\cite{Erben:2017hvr}.
The line is the KSRF relation with physical values for the $\rho$ mass and $f_{PS}$.
\label{fig:ksrf}}
\end{figure}

\begin{figure}
\begin{center}
\includegraphics[width=0.8\textwidth,clip]{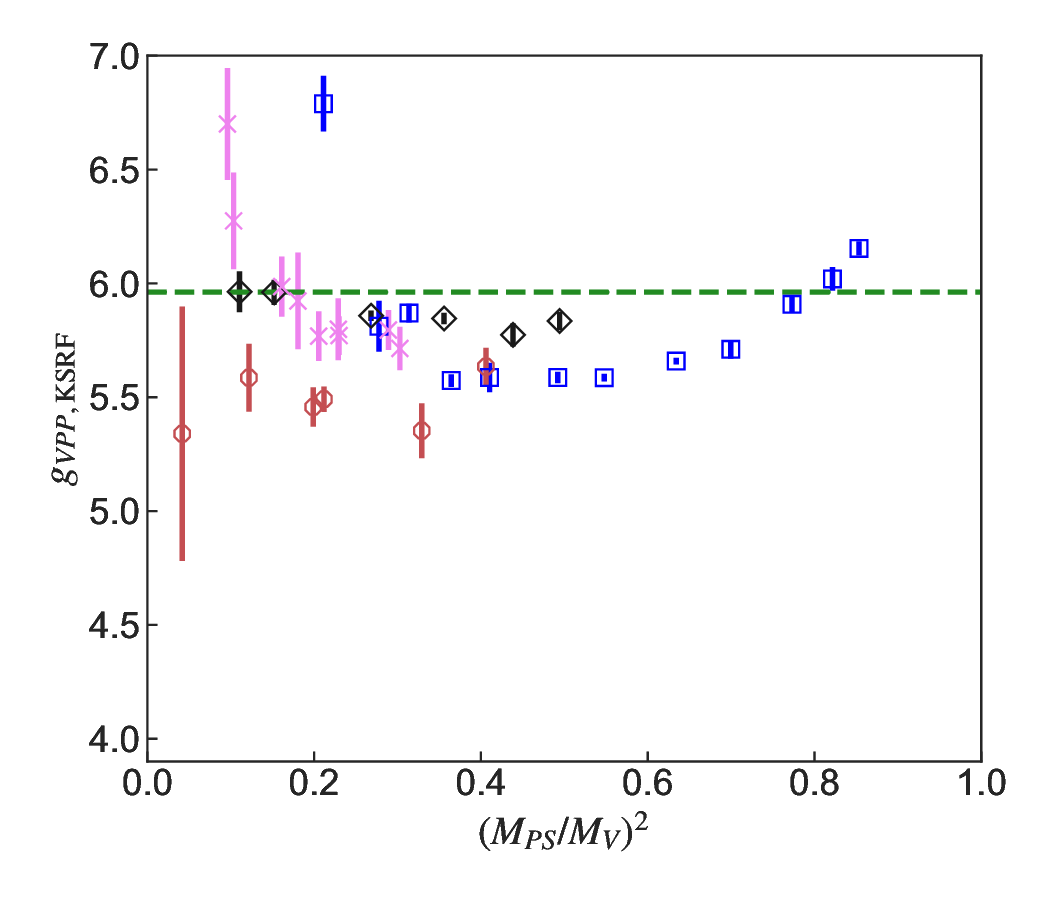}
\end{center}
\caption{Indirect determination of the vector meson decay constant $g_{VPP}$ from lattice calculations,
 as a function of $(M_{PS}/M_V)^2$, as inferred from the KSRF relation Eq.~(\ref{eq:ksrfv}).
 Data are black diamonds from Ref.~{\protect{\cite{WalkerLoud:2008bp}}},
red octagons from Ref.~{\protect{\cite{Aoki:2008sm}}},
violet crosses from Refs.~{\protect{\cite{Alexandrou:2008tn,Jansen:2009hr,Baron:2009wt}}},
blue squares from this work.  The star shows the physical $\rho$ meson decay constant.
\label{fig:ksrf2}}
\end{figure}

Lattice data for $g_{VPP}$ from several groups is displayed in Fig.~\ref{fig:ksrf}, along with
the KSRF relation itself, evaluated using the physical values of $M_V$ and $f_{PS}$.
The agreement of lattice data with the relation is excellent independent of the pion mass.  One may also
use the KSRF relation to indirectly estimate $g_{VPP}$ from lattice calculations of $M_V$ and $f_{PS}$; we show the result of this method in Fig.~\ref{fig:ksrf2}.  Although some significant deviations are seen from the direct results at lighter mass, the indirect approach is qualitatively accurate, giving $g_{VPP} \sim 6$ over a wide range of $(M_{PS}/M_V)$.

There are many lattice calculations of resonances which couple to two final state particles.
 Decays into three or more particles
is an active area of research.

\subsection{The $f_0$ or $\sigma$ meson \label{sec:sigma}}

In QCD, the $f_0$ meson is a broad scattering resonance; it is not a typical inclusion in calculations
 of the spectrum of light mesons because it must be carefully isolated using appropriate finite-volume 
scattering techniques if it is unstable.  However, at heavy quark masses the decay channel to two
 pseudoscalars is closed, and it can be studied using standard spectroscopy methods, although the 
presence of ``quark-line disconnected'' diagrams makes it computationally expensive to pursue. 
 Ref.~\cite{Kunihiro:2003yj} is an early study of the scalar meson on rather small volumes and extremely 
heavy quark masses, $(M_{PS}/M_V)^2 \sim 0.5-0.7$; they find $M_S > M_V$ in this entire range.

Lattice results for this state as a scattering resonance are beginning to appear: see Ref.~\cite{Briceno:2017qmb,Briceno:2016mjc,Guo:2018zss}.
Their data is at $(M_{PS}/M_V)^2= 0.1-0.2$ and the state ranges in mass from 460 to 760 MeV.
The $f_0$ is the lightest state in the hadron spectrum apart from the pions and as soon as it becomes heavier
than $2M_{PS}$ it becomes very broad.

There is a lattice literature on confining systems with a light scalar resonance, of the same order as $M_{PS}$.
 It is not QCD-like; instead, it appears in systems whose scale-dependent coupling constant runs very slowly as 
the energy scale varies.  This has led to arguments that the scalar is acting as a ``pseudo-dilaton'' in these
 systems whose lightness is associated with breaking of approximate scale invariance.  We will not say anything 
further about this interesting area of research here; the interested reader should consult
 Ref.~\cite{Witzel:2019jbe} for a recent review in the context of composite Higgs models, or the 
white paper Ref.~\cite{Brower:2019oor} for current and future prospects for lattice study and model 
understanding of an emergent light scalar meson.

\section{Conclusions \label{sec:conc}}

There is a wealth of lattice QCD data for SU(3) gauge theory at ``unphysical'' values of the quark
 mass parameters.  For the phenomenologist interested in hidden sectors or other models that contain
 an SU(3) gauge sector, we have attempted to gather and summarize a number of lattice results, and 
to elucidate how to interpret other lattice papers in a different context. 

For lattice QCD practitioners, we have a different remark: results at ``unphysical'' fermion masses
 may have an audience, and it may be useful to present them on  their own, rather than merely as
 intermediate steps on the way to the physical point of QCD.

Although we have focused on simpler quantities, there are substantial amounts of ``heavy QCD" lattice
 results available for nuclear physics, especially binding energies of small 
nuclei \cite{Beane:2012vq,Beane:2011iw,Beane:2013br,Orginos:2015aya,Sasaki:2017ysy}.  Some work has already been done in
 attempting to match these results on to nuclear EFTs \cite{Bansal:2017pwn}, which could provide
 a starting point for studying BSM scenarios in which the formation of BSM ``nuclei'' is of
 interest \cite{Krnjaic:2014xza,Gresham:2017zqi,Gresham:2017cvl,McDermott:2017vyk,Gresham:2018anj,Redi:2018muu}.  

There may be results which are not directly relevant for QCD, which have a place in 
phenomenology and could easily be generated.
An example of this would be spectroscopy and matrix elements of heavy fermion systems (i.e. quarkonia),
away from the charm and bottom quark masses. Perhaps appropriate data sets exist, but
neither we, nor the phenomenologists whose papers we have read, have noticed them.
Finally, we remark that it might not be too difficult to bring light hadron spectroscopy at
unphysical quark masses to the same level of precision as already exists for QCD
at the physical point.


\begin{acknowledgments}
We would like to thank
John Bulava,
Jim Halverson,
Robert McGehee,
and Yuhsin Tsai
 for conversations and correspondence.  We also thank Nick Evans and Suchita Kulkarni for drawing our attention to a labeling error in a previous version of this paper.  This material is based upon work supported by the U.S. Department of Energy, Office of Science, Office of High Energy Physics under Award Number DE-SC-0010005
 
\end{acknowledgments}

\appendix


\section{Normalization conventions and $f_{PS}$ \label{sec:fPS}}

The decay width of the pion is a physical quantity, in that it is an energy scale which is unambiguously determined by experimental measurement.  Although it is often used as one, the decay constant $f_{PS}$ is \emph{not} a physical quantity in the same sense; its precise value depends on various choices of convention.  The precise definition of these conventions is not always clear in the literature, so we will try to lay out some of the important details here.

The decay constant $f_{PS}$ itself is defined directly from the matrix element that describes the overlap of the pion field with the axial-vector current, relevant for e.g.~leptonic decays of the QCD pion:
\bee \label{eq:def_fPS}
\bra{0} J_A^{\mu}(0) \ket{\pi^+(p)} = i p^\mu \mathcal{N}_{PS} f_{PS}.
\ee
There are two ways in which different normalizing factors can enter into this equation and alter the numerical value of $f_{PS}$.  The first is that an arbitrary normalization $\mathcal{N}_{PS}$ may be included in the defining equation itself; this factor is usually set to 1, but the choice $\mathcal{N}_{PS} = \sqrt{2}$ sometimes occurs.  The second is that commonly, the decay constant will be defined with respect to the isospin eigenstate $\pi^a$ instead of the charged-pion state $\pi^+$, i.e.
\bee
\bra{0} J_A^{\mu,a}(0) \ket{\pi^b(p)} = ip^\mu \delta^{ab} \mathcal{N}_{PS} F_{PS}.
\ee
Since the charged pion states are defined as $\pi^\pm = (\pi^1 \pm i \pi^2) / \sqrt{2}$, this version of the decay constant is related to the one given above as $F_{PS} = f_{PS} / \sqrt{2}$.

Lattice calculations in particular almost exclusively work with the charged pion fields, so that (for example) in terms of quark content in QCD the current appearing in equation \ref{eq:def_fPS} is $\bar{d} \gamma^\mu \gamma^5 u$ and the pion field is $\bar{d} \gamma^5 u \ket{0}$.  In the results presented here we choose $\mathcal{N}_{PS} = 1$, which leads to the physical value $f_{PS} \approx 130$ MeV.

When in doubt, check a physical quantity like a decay width: for example, for any combination of conventions that lead to $f_{PS} \approx 130$ MeV in QCD, the result for tree-level standard model leptonic pion decay should be
\bee
\Gamma(\pi^+ \rightarrow \mu^+ \nu_\mu) = \frac{G_F^2 |V_{ud}|^2 f_{PS}^2}{8\pi} M_{PS} m_\mu^2 \left(1 - \frac{m_\mu^2}{M_{PS}^2} \right)^2.
\ee

Practical application of $f_{PS}$ often involves the use of chiral perturbation theory, where (at tree level and leading order) $f_{PS}$ may be identified\footnote{We will assume that $f_{PS} = F + ...$, but we caution that it is possible for additional normalizing factors to appear in the matching calculation to obtain the matrix element above from the chiral Lagrangian, so this relation should be confirmed for specific applications.} with the leading-order low-energy constant $F$.  Within the chiral Lagrangian, there are two normalization choices which can influence the numerical value of $F$.  The first is the normalization of the generators $T^a$ of the Lie algebra $\textrm{su}(N)$, which is typically expressed in terms of the trace normalization $\mathcal{N}_N$,
\bee
\Tr[ T^a T^b] = \mathcal{N}_N \delta^{ab}.
\ee
The second choice is the pion field normalization $\mathcal{N}_\pi$, which appears in the definition of the nonlinear matrix field $U$ as
\bee
U = \exp \left( \frac{i \mathcal{N}_\pi \pi^a T^a}{F} \right).
\ee
Requiring canonical normalization for the kinetic term of the $\pi^a$ fields fixes the kinetic term of the chiral Lagrangian to be
\bee
\mathcal{L}_{\rm kin} = \frac{F^2}{2\mathcal{N}_\pi^2 \mathcal{N}_N} \Tr \left[ \partial_\mu U \partial^\mu U^\dagger \right].
\ee
This relationship can be useful for inferring either $\mathcal{N}_\pi$ or $\mathcal{N}_N$ in cases where they are not clearly and explicitly stated.  Now, expanding the $U$ and $U^\dagger$ fields to second order in $\pi^a$ leads to a unique four-point interaction, schematically of the form
\bee
\mathcal{L}_{\rm kin} \sim \frac{1}{2} (\partial_\mu \pi)^2 + \frac{\mathcal{N}_\pi^2}{2F^2 \mathcal{N}_N} (\pi \partial_\mu \pi)^2 \Tr [ T^4 ] + ...
\ee
The coefficient of the four-point term must be independent of convention, because it directly determines the observable tree-level pion scattering amplitude.  We don't need to evaluate the trace $\Tr [T^4]$ directly, but we immediately see that it will scale as $\mathcal{N}_N^2$ if the normalization of the $T^a$ is changed.  Therefore, if we make two convention choices $A$ and $B$, we immediately find that the values of $F$ are related as
\bee
\frac{F_B^2}{F_A^2} = \frac{(\mathcal{N}_\pi^2 \mathcal{N}_N)_B}{(\mathcal{N}_\pi^2 \mathcal{N}_N)_A}.
\ee
All that remains is to fix a fiducial numerical value of $F$ for a specific choice of conventions: in principle, this requires a matching calculation onto a physical process such as leptonic pion decay.  We adopt the results of Scherer and Schindler\cite{Scherer:2005ri}, who find $F = 93$ MeV with the choices $\mathcal{N}_\pi = 1$ and $\mathcal{N}_N = 2$, leading to the general result.
\bee
F = 93\ \textrm{MeV} \times \mathcal{N}_\pi \times \sqrt{\frac{\mathcal{N}_N}{2}}.
\ee

\section{Lattice details for $f_V$ and $f_A$}
The data sets from which our values of $f_V$ and especially $f_A$ were extracted were
extensions of the ones published in Ref.~\cite{DeGrand:2016pur}.  The simulations used 
the Wilson gauge action and clover fermions with
normalized hypercubic links~\cite{Hasenfratz:2001hp,Hasenfratz:2007rf}.
They had $N_f=2$ flavors of degenerate mass fermions.
All lattice volumes are $16^3\times 32$. One value of bare gauge coupling was simulated.
Ref.~\cite{DeGrand:2016pur} used the shorter version of the Sommer parameter to set the scale
and so we continue to do that here.

Calculations of the axial vector meson require high statistics since the signal is noisy. Our data sets consist of 400
lattices (except for $\kappa=0.127$, with 203 lattices), spaced ten molecular dynamics time steps apart.

The raw lattice numbers for the decay constants are converted into continuum regularization.
via the old tadpole-improved procedure of Lepage and Mackenzie
\cite{Lepage:1992xa}. We work at one loop.
In this scheme a continuum-regulated fermionic bilinear quantity $F$ with engineering
dimension $D$ (for example the
$\overline{MS}$ (modified minimal subtraction) value at scale $\mu$) is related to the lattice value by
\bee
F(\mu) = a^D F(a) (1 - \frac{3 \kappa}{4 \kappa_c}) Z_Q
\ee
and at scale $\mu a=1$,
\bee
Z_Q= 1 + \alpha \frac{C_F}{4\pi} z_Q
\ee
where $\alpha=g^2/(4\pi)$, $C_F$ is the usual quadratic Casimir,  and $z_Q$ is
 a scheme matching number.
(Ours are tabulated in  Ref.~\cite{DeGrand:2002vu}.) The axial vector and
vector Z-factors
 are only a few percent different from unity for nHYP clover fermions and so $Z_Q$ is taken to be unity.
$\kappa_c$ is the value of the hopping parameter where  the axial Ward identity (AWI) quark mass and the pion
 mass vanishes.
The lattice spacing depends on the bare simulation parameters, so
we determined the values of $\kappa_c$ by fitting the dimensionless combination
 $r_1 m_q$ to a linear dependence on $\kappa$.

We present our results in two tables. The first is intended for lattice practitioners: it shows (in Table
\ref{tab:lattice} the hopping parameter, Sommer radius, AWI quark mass, pseudoscalar, 
vector, and axial vector masses, and the lattice pseudoscalar, vector and axial vector  decay constants.

The second table (Table \ref{tab:contin}
 is more useful for phenomenologists: it shows the squared pseudoscalar to vector mass ratio,
the vector and axial masses in MeV, and the three decay constants (MeV for $f_{PS}$) in continuum regularization.
These are the quantities plotted in Figs.~\ref{fig:meson}, \ref{fig:fpi} and \ref{fig:decaynewpirho}.

\section{Lattice datasets}

Here we tabulate the new SU(3) lattice results obtained for the purposes of this paper.

\clearpage

\begin{table}
\begin{tabular}{c c c c c c c c c }
\hline
$\beta=5.4$  & $\kappa_c=0.12838$ & & & & &   \\
\hline
$\kappa$ & $r_1/a$  & $a\,m_q$ & $a\,M_{PS}$ & $a\,M_V$  & $aM_A$ & $af_{PS, {\rm bare}}$ & $f_{V, \rm{bare}}$ & $f_{A, \rm{bare}}$ \\
\hline
 0.1180 &  2.272(27) &  0.3230 &  1.127(2) &  1.220(2) &  1.535(8) &  0.638(3) &  0.670(4) &  0.365(3) \\
 0.1200 &  2.352(14) &  0.2620 &  0.997(2) &  1.100(3) &  1.415(8) &  0.611(5) &  0.726(4) &  0.431(5) \\
 0.1220 &  2.552(20) &  0.2010 &  0.830(1) &  0.944(2) &  1.251(9) &  0.556(3) &  0.787(3) &  0.466(7) \\
 0.1240 &  2.736(28) &  0.1390 &  0.665(3) &  0.795(3) &  1.063(7) &  0.505(3) &   0.875(16) &  0.524(7) \\
 0.1250 &  2.950(20) &  0.1070 &  0.563(1) &  0.707(1) &  0.979(3) &  0.463(1) &   0.932(12) &  0.576(5) \\
 0.1260 &  3.080(30) &  0.0730 &  0.453(1) &  0.612(1) &  0.874(4) &  0.415(1) &  0.973(7) &  0.629(4) \\
 0.1265 &  3.110(30) &  0.0580 &  0.395(1) &  0.563(2) &  0.806(7) &  0.386(1) &  0.996(5) &  0.657(4) \\
 0.1270 &  3.230(30) &  0.0410 &  0.328(1) &  0.512(4) &  0.738(9) &  0.355(3) &  1.052(8) &  0.714(5) \\
 0.1272 &  3.300(30) &  0.0340 &  0.300(1) &  0.497(2) &  0.718(5) &  0.347(1) &  1.089(5) &   0.685(14) \\
 0.1274 &  3.320(20) &  0.0274 &  0.270(1) &  0.482(3) &  0.695(4) &  0.321(1) &  1.088(5) &  0.737(6) \\
 0.1276 &  3.460(20) &  0.0206 &  0.234(2) &  0.444(8) &   0.656(10) &  0.300(2) &   1.073(11) &  0.719(7) \\
 0.1278 &  3.410(30) &  0.0140 &  0.204(2) &  0.444(4) &  0.626(5) &  0.258(4) &  1.115(8) &  0.798(9) \\
\hline
 \end{tabular}
\caption{ Masses in lattice units for the $SU(3)$ data sets. From left to right, the entries
are the  hopping parameter $\kappa$, the relative scale $r_1/a$, the Axial Ward Identity quark mass,
 the pseudoscalar mass, the  vector meson mass, the axial vector meson mass, the
pseudoscalar decay constant, the vector meson decay constant, and the axial vector meson decay constant.  Decay constants are not renormalized.
\label{tab:lattice}}
\end{table}

\begin{table}
\begin{tabular}{c c c c c c }
\hline
$(M_{PS}/M_V)^2$ & $M_{V}$, MeV  & $M_A$, MeV &$f_{PS}$, MeV & $f_V$ & $f_A$ \\
\hline
0.8534 &     1745(21) &  2195(28) &  283.5(3.6) &   0.2054(12) &  0.11200(92) \\
 0.8215 &     1628(11) &  2095(17) &  270.5(2.7) &   0.2142(12) &   0.1273(15) \\
 0.7731 &     1516(12) &  2009(21) &  256.6(2.4) &  0.22316(85) &   0.1322(20) \\
 0.6997 &     1369(15) &  1831(22) &  239.7(2.8) &   0.2380(44) &   0.1426(19) \\
 0.6341 &     1313(9) &  1818(14) &  232.0(1.7) &   0.2481(32) &   0.1535(13) \\
 0.5479 &     1186(12) &  1694(18) &  212.4(2.1) &   0.2535(18) &   0.1640(10) \\
 0.4922 &     1102(11) &  1578(20) &  197.3(2.0) &   0.2566(13) &   0.1694(10) \\
 0.4104 &     1041(13) &  1500(23) &  186.3(2.3) &   0.2680(20) &   0.1820(13) \\
 0.3644 &     1032(10) &  1491(17) &  185.2(1.8) &   0.2761(13) &   0.1738(36) \\
 0.3138 &     1007(9) &  1452(12) &  171.6(1.2) &   0.2746(13) &   0.1862(15) \\
 0.2778 &      967(18) &  1429(23) &  166.4(1.5) &   0.2696(28) &   0.1808(18) \\
 0.2111 &      953(12) &  1344(16) &  140.4(2.5) &   0.2789(20) &   0.1997(23) \\
\hline
 \end{tabular}
\caption{ Continuum results. From left to right, the entries
are the squared ratio of pseudoscalar to vector masses,
 the  vector meson mass in MeV, the axial vector meson mass in  MeV, the
pseudoscalar decay constant in MeV, the vector meson decay constant, and the axial vector meson decay constant.  All decay constants are renormalized as described above.
\label{tab:contin}}
\end{table}


\vfill\eject

\end{document}